\documentclass[a4paper,fleqn,usenatbib]{mnras}

\usepackage{natbib}
\usepackage{amsmath,amssymb}
\usepackage{epsfig,graphicx}
\usepackage{subcaption}
\usepackage{float}
\usepackage{lastpage}

\DeclareGraphicsExtensions{.pdf,.png,.jpg,.eps}
\newcommand{\wgg}{\ensuremath{w_{gg}}}
\newcommand{\wgp}{\ensuremath{w_{g+}}}

\newcommand{\wpp}{\ensuremath{w_{++}}}

\newcommand{\xigg}{\ensuremath{\xi_{gg}}}
\newcommand{\xigp}{\ensuremath{\xi_{g+}}}
\newcommand{\xipp}{\ensuremath{\xi_{++}}}

\newcommand{\mpch}{\ensuremath{h^{-1}\text{Mpc}}}

\def\reff@jnl#1{{\rm#1\/}}

\def\aj{\reff@jnl{AJ}}                  
\def\araa{\reff@jnl{ARA\&A}}            
\def\apj{\reff@jnl{ApJ}}                        
\def\apjl{\reff@jnl{ApJ}}               
\def\apjs{\reff@jnl{ApJS}}              
\def\apss{\reff@jnl{Ap\&SS}}            
\def\aap{\reff@jnl{A\&A}}               
\def\aapr{\reff@jnl{A\&A~Rev.}}         
\def\aaps{\reff@jnl{A\&AS}}             
\def\baas{\reff@jnl{BAAS}}              
\def\jrasc{\reff@jnl{JRASC}}            
\def\memras{\reff@jnl{MmRAS}}           
\def\mnras{\reff@jnl{MNRAS}}            
\def\physrep{\reff@jnl{Phys.Rep.}}
\def\pra{\reff@jnl{Phys.Rev.A}}         
\def\prb{\reff@jnl{Phys.Rev.B}}         
\def\prc{\reff@jnl{Phys.Rev.C}}         
\def\prd{\reff@jnl{Phys.Rev.D}}         
\def\prl{\reff@jnl{Phys.Rev.Lett}}      
\def\pasp{\reff@jnl{PASP}}              
\def\pasj{\reff@jnl{PASJ}}              
\def\skytel{\reff@jnl{S\&T}}            
\def\solphys{\reff@jnl{Solar~Phys.}}    
\def\sovast{\reff@jnl{Soviet~Ast.}}     
\def\ssr{\reff@jnl{Space~Sci.Rev.}}     
\def\nat{\reff@jnl{Nature}}             

\newcommand{\hmpc}{\ensuremath{h^{-1}\mathrm{Mpc}}}

\newcommand{\beq}{\begin{equation}}
\newcommand{\eeq}{\end{equation}}
\newcommand{\beqa}{\begin{eqnarray}}
\newcommand{\eeqa}{\end{eqnarray}}

\newcommand{\erms}{\ensuremath{e_\mathrm{rms}}}

\usepackage{epstopdf}

\graphicspath{{figures/}}

\newcommand{\referee}[1]{{{#1}}}

\newcommand{\pI}{{Paper I}}
\newcommand{\dev}{{de Vaucouleurs}}
\newcommand{\apsf}{\ensuremath{A_{\text{PSF}}}}

\title[Intrinsic Alignments in BOSS-II]{Intrinsic alignments of BOSS LOWZ galaxies II: Impact of
  shape measurement methods}

\author[Singh \& Mandelbaum]{ Sukhdeep Singh$^1$\thanks{\tt
    sukhdeep@cmu.edu}, Rachel Mandelbaum$^{1}$ \\ $^1$McWilliams
  Center for Cosmology, Department of Physics, Carnegie Mellon
  University, Pittsburgh, PA 15213, USA\\ } \date{\today}
\date{Accepted XXX. Received YYY; in original form ZZZ}
\pubyear{2015}
\begin{document}
\label{firstpage}
\pagerange{\pageref{firstpage}--\pageref{LastPage}}
\maketitle
\begin{abstract}
Measurements of intrinsic alignments  of galaxy shapes with the large-scale density field, and the
inferred intrinsic alignments model parameters, are sensitive to the shape measurement
methods used. In this paper we measure the intrinsic alignments of the Sloan Digital Sky Survey-III
(SDSS-III) Baryon Oscillation Spectroscopic Survey (BOSS) LOWZ galaxies using three different shape 
measurement methods (re-Gaussianization, isophotal, and \dev), identifying a variation in the 
inferred intrinsic alignments amplitude at the 40\%
level between these methods, independent of the galaxy luminosity or other
properties.   We also carry out a suite of systematics tests on the shapes and
their two-point correlation functions, identifying a pronounced contribution from additive PSF systematics in
the \dev\ shapes.  Since different methods 
measure galaxy shapes at different effective radii, the trends we identify in the intrinsic
alignments amplitude are consistent with the interpretation that the outer 
regions of galaxy shapes are more responsive to tidal fields, resulting in isophote twisting and
stronger alignments for isophotal shapes. We observe environment dependence of ellipticity, with brightest galaxies in groups being rounder on average compared to satellite and field galaxies.
We also study the anisotropy in intrinsic alignments measurements introduced by projected shapes,
finding effects consistent with predictions of the nonlinear alignment model and hydrodynamic
simulations.  The large variations seen using the different shape measurement methods 
 have important implications for intrinsic alignments forecasting and
mitigation with future surveys.
\end{abstract}

\begin{keywords}
  galaxies: evolution\ --- cosmology: observations
  --- large-scale structure of Universe\ --- gravitational
  lensing: weak
\end{keywords}

\section{Introduction}\label{sec:intro}{
\vspace{1in}

Weak gravitational lensing \citep[for a review, see][]{2010RPPh...73h6901M,2013PhR...530...87W}, the deflection of light from distant
objects by mass in more nearby 
foregrounds, results in coherent distortions of galaxy shapes that are measured statistically, by
averaging over large ensembles of galaxies.  It has the power to reveal the dark matter halos in
which galaxies and galaxy clusters reside \citep[e.g.,][]{2014MNRAS.437.2111V,2015MNRAS.449.1352C,2015MNRAS.446.1356H,2015MNRAS.447..298H,2015arXiv150502781Z}, to constrain the growth of cosmic
structure and thus the nature of dark energy \citep[e.g.,][]{2013MNRAS.432.2433H,2013ApJ...765...74J,2013MNRAS.432.1544M}, and even to constrain the theory of
gravity on cosmological scales \citep[e.g.,][]{2010Natur.464..256R,2013MNRAS.429.2249S,2015MNRAS.449.4326P}.
Intrinsic alignments of galaxies (IA; for a review, see \citealt{2015arXiv150405456J,Troxel2015,2015arXiv150405465K,2015arXiv150405546K}), 
the coherent alignment of galaxy shapes with each other 
(II) or with the local density field (GI), result in a violation of the assumption that galaxy shapes are not
intrinsically correlated and that any  observed shape correlations are from gravitational lensing.
Thus, intrinsic alignments are 
an important astrophysical systematic for weak lensing surveys.
 
 \cite{Singh2015} (hereafter \pI) studied the IA of galaxies in one of the Sloan Digital Sky Survey-III
 (SDSS-III) Baryon Oscillation Spectroscopic Survey (BOSS) galaxy samples, called LOWZ.  This sample
 consists of Luminous Red Galaxies (LRGs) with a comoving number density of $\sim 3\times 10^{-4}
 (h/\text{Mpc})^3$ from $0.16<z<0.36$.  The large sample size and high signal-to-noise ratio enabled
 not only a detection of IA, but a study of 
its dependence on galaxy properties such as mass, luminosity and 
environment, finding strong correlations of IA with the host halo mass and galaxy luminosity 
\citep[see also][]{Joachimi2011}. A commonly-adopted theoretical model called the linear alignment
model \citep[][]{Catelan2001,Hirata2003}, which relates
the galaxy alignments to the tidal field from large-scale structure at the time of galaxy formation,
was found to provide a good description of the data for projected separation $r_p>6~h^{-1}$Mpc, provided that the non-linear matter power
spectrum was used 
\citep[NLA model;][]{Bridle2007}.

A natural goal of such studies is to predict the intrinsic alignment contamination in weak lensing
surveys, and to provide templates that can be used to marginalize over this effect.  However, 
different IA studies in the literature
\citep[e.g.,][]{Mandelbaum2006,Hirata2007,Okumura2009,Joachimi2011,Blazek2011,Hao2011,Hung2012,Li2013,Chisari2014,Sifon2015,Singh2015}
use different galaxy shape measurement methods and ensemble IA estimators, which makes it difficult
to compare them or to combine their results into a single comprehensive view of the subject.  
For example, \cite{Okumura2009} used isophotal 
shape measurements from SDSS data release 7 (DR7) to measure shape-shape correlations for the SDSS LRG sample at high 
significance which were then interpreted by \cite{Blazek2011} in the context of the NLA model. However, in \pI\, using 
re-Gaussianization shapes and a larger BOSS low redshift sample (LOWZ-DR11) from SDSS-III, we measured 
the shape-shape correlation function to be consistent with zero. Though 
the two measurements were shown to be statistically consistent in \pI, the reason for the varying
detection significance in the two studies is not well understood, and worth investigation.

IA measurements with different shape measurements may be particularly difficult to compare due to different
systematic errors and ranges of galaxy radius probed by different methods. 
First,  galaxy shape measurements are affected by observational systematics such as the point-spread
function (PSF), pixel noise, errors in estimation and subtraction of sky level, and blending with the
light profiles of nearby galaxies. Different shape 
measurement methods treat these systematics differently or in some cases ignore them. Improper
treatment of systematic effects in galaxy shapes can propagate into IA measurements. For example,
\cite{Hao2011} found that  using 
isophotal shapes from the SDSS to measure the alignments of satellite shapes around brightest
cluster galaxies (BCGs) that these alignments correlated with the 
{\em apparent} magnitude of the BCG. This result was interpreted as a systematic error in isophotal shapes of satellite galaxies from  
BCG light leaking into the satellite shapes, an effect that is more complicated than a 
multiplicative bias. 

Also, different shape measurement methods use different radial weight functions and thus probe
galaxy shapes at 
different effective radii.  Intrinsic variation in the galaxy shapes with radius can thus affect the
IA measurements.  These could be gradients in the ellipticity with radius (with galaxies being
intrinsically more or less round in the outer regions), or isophotal twisting due to the outer parts
of galaxies being {more aligned with external tidal field than the inner parts. } 
For example, \cite{Tenneti2015b} found in hydrodynamic simulations 
that galaxies become rounder in their outer regions. 
\cite{Velliscig2015}, on the other hand, found that the projected RMS ellipticity,
  $e_\text{rms}$, is lower when measured using star particles within the half-light radius compared to using all star particles {\citep[see also][]{Chisari2015}}.
In observations, the magnitude and sign of ellipticity gradients 
have been found to be correlated with the galaxy environment \citep{Tullio1978,Tullio1979,Pasquali2006}. 
{Ellipticity variations using different shape measurements will affect 
inferences made using ensemble IA 
estimators that include the ellipticity rather than just the position angle 
\citep[e.g.,][]{Mandelbaum2006,Hirata2007,Joachimi2011,Blazek2011,Singh2015}.
}

Several studies have also detected isophote twists in small samples of elliptical galaxies, whereby the position angle in 
the measured galaxy shape changes with radius
\citep{Wyatt1953,Abramenko1978,Kormendy1982,Fasano1989,Nieto1992,Lauer2005}. The isophote twisting may 
originate from varying triaxiality of galaxies with radii \citep[see, e.g.,][]{Romanowsky1998} though 
\cite{Kormendy1982} pointed out that the outer regions of galaxies are more susceptible to tidal fields, which can 
result in isophote twisting.  In simulations, \cite{Kuhlen2007} observed the effects of shape twisting from tidal 
interaction when measuring the radial alignments of dark matter subhalos. The radial alignment signal of subhaloes 
increased monotonically with the radius at which the subhalo shape was measured. 
If such results also apply to galaxies, then a stronger IA signal from shape measurements that 
probe the outer regions of galaxies would be expected. In support of this inference, 
\cite{Velliscig2015b} found using hydrodynamic simulations that using star particles
  within the half-light radius to define the galaxy shapes results in 
lower IA signal compared to using all the star particles {\citep[see also ][but note
  that that work attributes the differences to ellipticity variations rather than to isophote twists]{Chisari2015}}.

Finally, we measure only the projected shapes of galaxies, which are insensitive to line-of-sight
galaxy alignments. This effect introduces anisotropy in the redshift-space structure of
IA\footnote{{This anisotropy due to projected shapes is also present  in real space, but we
    will use the term redshift-space throughout this paper since measurements are made in redshift-space.}}, and
different estimators of IA vary in their sensitivity to this anisotropy. For example, the
redshift-space structure of the 3D shape-density cross-correlation function, \xigp, includes the
redshift-space structure of the 3D galaxy-galaxy auto-correlation function, \xigg; however, 
the mean IA shear, $\langle\gamma\rangle$, is independent of \xigg\ {($\xigp=\langle\gamma\rangle(1+\xigg)$, \citealt{Blazek2015})}. Depending on the relative
importance of variations in IA and galaxy clustering, \xigp\ and $\langle\gamma\rangle$ may have
different redshift-space structure.  
The variations in the sensitivity to redshift-space structure between different IA estimators can
complicate a quantitive comparisons between studies using different estimators. 

In this paper, we repeat the analysis of \pI\ using three different shape measurement methods, and
carry out numerous systematics tests, to study the radial and environment dependence of galaxy
shapes and of the IA signal. 
We also use the methodology of \cite{Blazek2011} to understand the origin of differences in their 
{shape-shape correlations (\wpp)} measurement and the one in \pI. 
Finally, to understand the redshift-space structure of intrinsic alignments, we investigate the IA
signals as a function of projected and line of sight separations and compare the results with NLA model predictions.

Throughout we use a standard flat $\Lambda$CDM cosmology with $h=0.7$, $\Omega_b=0.046$,
$\Omega_{DM}=0.236$, $\Omega_\Lambda=0.718$, $n_s=0.9646$, $\sigma_8=0.817$
\citep[WMAP9,][]{Hinshaw2013}. All distances are in comoving \mpch, though $h=0.7$ 
was used to calculate absolute magnitudes and to generate predictions for the matter power
spectrum.  }

\section{Formalism and Methodology}

	Details of intrinsic alignments models and correlation estimators are given in \pI. In this section, we 
	briefly summarize the important points.

	\subsection{The nonlinear alignment (NLA) model}
		The linear alignment (LA) model predicts that IA are set at the time of galaxy formation 
		\citep{Catelan2001}, 
		with galaxy shapes being aligned with the tidal fields present during galaxy 
		formation.   
		This assumption allows us to write intrinsic shear in terms of primordial potential $\phi_p$
		\begin{equation}
			\gamma^I=(\gamma^I_+, \gamma^I_\times)=-\frac{C_1}{4\pi G}(\partial^2_x-
			\partial^2_y,\partial_x\partial_y)\phi_p,
			\label{eqn:gamma_phi}
		\end{equation}
		with alignment strength defined by an amplitude parameter $C_1$. In our sign convention, positive (negative) 
		$\gamma^I_+$ indicates alignments along (perpendicular to) the direction of the tidal field while positive 
		(negative) 
		$\gamma^I_\times$ indicates alignments along the direction at $45$ ($135$) degrees from the
        direction of the tidal field.
		Assuming a linear galaxy bias relating matter overdensities $\delta_m$ and galaxy densities
        $\delta_g=b~\delta_m$, the power spectrum of galaxy-shape and shape-shape correlations can be written as 
		\citep{Hirata2003} 
		\begin{align}
			P_{g+}(\vec{k},z)&=A_I b \frac{C_1\rho_{\text{crit}}\Omega_m}{D(z)} \frac{k_x^2-
			k_y^2}{k^2} P_\delta^\text{lin} (\vec{k},z)\label{eqn:LA+}\\
			P_{++}(\vec{k},z)&=f_{II}\left(A_I \frac{C_1\rho_{\text{crit}}\Omega_m}{D(z)} 
			\frac{k_x^2-k_y^2}{k^2} \right)^2P_\delta^\text{lin} (\vec{k},z)\label{eqn:LA++}\\
			P_{g\times}(\vec{k},z)&=A_I b\frac{C_1\rho_{\text{crit}}\Omega_m}{D(z)}\frac{k_x 
			k_y}{k^2}P_\delta^\text{lin}(\vec{k},z)\label{eqn:LAx}
		\end{align} 
		$P_\delta^\text{lin}$ is the linear matter power spectrum. $P_{g+}$ ($P_{g\times}$) is the
        cross-power spectrum between the galaxy density field and the 
		shear component along (at $45^\circ$ from) the line joining the galaxy pair. $P_{++}$ 
		is the shape-shape correlation with shear component along the line joining the galaxy
        pair. Following \cite{Joachimi2011}, we fix $C_1\rho_\text{crit}=0.0134$ and use {a dimensionless} 
        constant $A_I$ to measure the IA amplitude, which
        is primarily constrained from density-shape 
		cross-correlations. To allow for departure from the LA model, we have 
		introduced an additional free parameter $f_{II}$, which would be 1 in the case that
	         the LA model is correct. In the NLA model, to extend the linear 
		alignment model to the non-linear regime, the 
		linear matter power spectrum is replaced with the non-linear matter power spectrum
        \citep{Bridle2007}   using an updated 
		halo-fit model \citep{Smith2003,Takahashi2012}.
	As shown by \cite{Blazek2015}, the NLA model
        neglects other terms that are important at the same order; however, in \pI\ we found it
        provided an adequate fit to our intrinsic alignments measurements down to $\sim 6$~\hmpc.
		
		The power spectra in Eqs.~\eqref{eqn:LA+}--\eqref{eqn:LAx} can be Fourier 
		transformed to obtain the 3D correlation functions as a function of comoving projected
        separation $r_p$ and line-of-sight separation $\Pi$,
		\begin{align}
			\xi_{AB}(r_p,\Pi,z)=&\int \frac{\mathrm{d}^2k_\perp\mathrm{d}k_z}{(2\pi)^3}P_{AB}
			(\vec{k},z)\left(1+\beta_A\mu^2\right)\nonumber \\ &\left(1+\beta_B
			\mu^2\right)e^{i(r_p.k_\perp+\Pi k_z)}.\label{eqn:xi}
		\end{align}
		The Kaiser factor ($1+\beta\mu^2$) accounts for the effect of linear redshift-space
        distortions (RSD; \citealt{Kaiser1987}). 
		As shown in \pI, for power spectra that include intrinsic shapes, $\beta_{+,\times}=0$
        {(the shear field is not affected by RSD to first order)}. For
        galaxies, $\beta(z)=f(z)/b$, where the linear growth factor $f(z)\sim \Omega_m(z)^{0.55}$ in
        the 
		$\Lambda$CDM model. 
		
		In the data, we measure the projected correlation function, $w_{AB}(r_p)$, which can be obtained by
		projecting the 3D correlation function in Eq.~\eqref{eqn:xi} along the line-of-sight separation:
		\begin{align}
		w_{AB}(r_p)=&\int\mathrm{d}z \,W(z) \int\mathrm{d}\Pi\,\xi_{AB}(r_p,\Pi,z).\label{eqn:w}
		\end{align}
		$W(z)$ is the redshift window function \citep{Mandelbaum2011}
		\begin{equation}
			W(z)=\frac{p_A(z)p_B(z)}{\chi^2 (z)\mathrm{d}\chi/\mathrm{d}z} \left[\int 
			\frac{p_A(z)p_B(z)}{\chi^2 (z)\mathrm{d}\chi/\mathrm{d}z} \mathrm{d}z\right]^{-1}
		\end{equation}
		
		Assuming cross-correlation between samples of galaxies with shapes S and others that trace
        the density field D with in principle different biases $b_S$ and $b_D$, the different projected correlation functions are
		\begin{align}
			w_{gg}(r_p)=\frac{b_S b_D}{\pi^2}\int\mathrm{d}z \,{W(z)}\int_0^{\infty}\mathrm{d}
			k_z\int_0^{\infty}\mathrm{d}k_{\perp}\frac{k_\perp}{k_z} P(\vec{k},z)\nonumber\\
			\sin(k_z\Pi_\text{max})J_0(k_\perp r_p)\left(1+\beta_S\mu^2\right)\left(1+\beta_D
			\mu^2\right)\label{eqn:wgg}
		\end{align}
		\begin{align}
			w_{g+}(r_p)=\frac{A_I b_D C_1 \rho_{\text{crit}} \Omega_m}{\pi^2}\int\mathrm{d}z 
			\frac{W(z)}{D(z)}\int_0^{\infty}\mathrm{d}k_z\int_0^{\infty} \mathrm{d}k_{\perp}\nonumber\\
			\frac{k_\perp^3}{(k_\perp^2+k_z^2)k_z} P(\vec{k},z)\sin(k_z\Pi_\text{max})J_2(k_
			\perp r_p) \left(1+\beta_{D}\mu^2\right) \label{eqn:wgp}
		\end{align}
		\begin{align}
			&w_{++}(r_p)=f_{II}\frac{\left(A_I C_1 \rho_{\text{crit}} \Omega_m\right)^2}{2\pi^2}\int
			\mathrm{d}z \frac{W(z)}{D(z)^2}\int_0^{\infty}\mathrm{d}k_z\int_0^{\infty} 
			\mathrm{d}k_{\perp}\nonumber\\ &\frac{k_\perp^5}{(k_\perp^2+k_z^2)^2k_z} P(\vec{k},z)\sin(k_z\Pi_
			\text{max})
			\times [J_0(k_\perp r_p)+J_4(k_\perp r_p)] \label{eqn:wpp}
		\end{align}
		Whether we fit the three correlation functions (\wgg, \wgp, and \wpp) jointly or
        independently to this model, 
		the galaxy bias, intrinsic alignment amplitude $A_I$, and $f_{II}$ are primarily constrained by \wgg, 
		\wgp, and \wpp, respectively. This is due to the different sensitivities of these signals to the three
        parameters, along with the different signal-to-noise ratio ($S/N$) in the 
		measurements. The $S/N$ degrades with increasing factors of galaxy shape due to the shape noise
        they add to 
		the correlation function.
		
	\subsection{Correlation function estimators}
We calculate cross-correlation functions between sets of galaxies for which we wish to estimate the
intrinsic alignments (the ``shape sample'') and those used to trace the density field (the ``density
sample'').  
		We use a generalized Landy-Szalay estimator \citep{Landy1993} to compute the correlation 
		functions:
			\begin{align}	
				\xi_{gg}=\frac{(S-R_S)(D-R_D)}{R_SR_D}&=\frac{SD-R_SD-SR_D}
				{R_SR_D}+1 \notag
			\end{align}
			\begin{align}	
				\xi_{g+}&=\frac{S_+D-S_+R_D}{R_SR_D}\notag
			\end{align}
			\begin{align}	
				\xi_{++}&=\frac{S_+S_+}{R_SR_S},\notag
			\end{align}
			where $S$ and $D$ represent the galaxy counts in the shape and density samples, and 
			$R_S$ and $R_D$ are sets of random points corresponding to these samples. The terms
            involving shears for the galaxies are  
			\begin{align}	
				S_+X&=\sum_{i\in S, j\in X}\gamma_+^{(i)}(j|i),\notag\\
				S_+S_+&=\sum_{i\in S, j\in S}\gamma_+^{(i)}(i|j)\gamma_+^{(j)}(j|i)\notag.
			\end{align}
			 Here $\gamma_{+}^{(i)}(j|i)$ represents the component of the shear for galaxy $i$ along
             the line joining it to galaxy $j$. Positive (negative) $\gamma_+$ implies radial
             (tangential) alignments. 

			Measuring projected correlations involves summation over bins in $\Pi$, 
			\begin{equation}	
				w_{AB}=\int^{\Pi_{\text{max}}}_{-\Pi_{\text{max}}}\xi_{AB}(r_p, \Pi)\,\mathrm{d}\Pi.
			\end{equation}
			We use $\Pi_{\text{max}}=100\mpch$ and $d\Pi=10\mpch$.
			
			To calculate the covariance matrices of the $w_{AB}$, we divide the sample into 100
            equal-area regions, compute the 
			signal by excluding one region at a time and then compute the jackknife variance from 
			the 100 jackknife samples. See \pI\ for more details.
	
	\subsection{Anisotropy}\label{ssec:anisotropy}

		It is well known that redshift-space distortions (RSD) introduce anisotropy in the observed 
		galaxy 
		clustering in redshift space \citep[e.g.,][]{Kaiser1987,Beutler2014}. In \pI, it was shown that the IA 
		measurements are not affect by RSD to first order. However, IA are affected by another 
		source of anisotropy, the projected shapes of galaxies, due to which we cannot measure 
		the line-of-sight IA signal. Hence, the IA signal is expected to fall faster with increasing
        line-of-sight separation, $\Pi$, than with $r_p$.
		
		To understand this anisotropy, we study the IA signal in $(r_p,\Pi)$ space and 
		compare it with the NLA model prediction. To compute the model predictions, we compute the 
		real part of $\xi_{AB}$ as follows:
		\begin{align}
			\xi_{gg}(r_p,\Pi)=&\frac{b_S b_D}{2\pi^2}\int\mathrm{d}z \,{W(z)}\int_0^{\infty}
			\mathrm{d}
			k_z\cos(k_z\Pi)\nonumber \\ &\int_0^{\infty}\mathrm{d}k_{\perp}\frac{k_\perp}{k_z} 
			P(\vec{k},z)
			J_0(k_\perp r_p)\left(1+\beta_S\mu^2\right)\nonumber \\ &\left(1+\beta_D
			\mu^2\right)\label{eqn:xigg}\\  
			\xi_{g+}(r_p,\Pi)=&\frac{A_I b_D C_1 \rho_{\text{crit}} \Omega_m}{2\pi^2}\int
			\mathrm{d}z 
			\frac{W(z)}{D(z)}\int_0^{\infty}\mathrm{d}k_z\cos(k_z\Pi)\nonumber \\ &\int_0^{\infty}
			\mathrm{d}k_{\perp}
			\frac{k_\perp^3}{(k_\perp^2+k_z^2)k_z} P(\vec{k},z)J_2(k_
			\perp r_p) \nonumber \\ 
			&\left(1+\beta_{D}\mu^2\right) \label{eqn:xigp}\\
			\xi_{++}(r_p,\Pi)=&f_{II}\frac{\left(A_I C_1 \rho_{\text{crit}} \Omega_m\right)^2}
			{4\pi^2}\int
			\mathrm{d}z \frac{W(z)}{D(z)^2}\int_0^{\infty}\mathrm{d}k_z\nonumber \\ &\cos(k_z
			\Pi)\int_0^{\infty} \mathrm{d}
			k_{\perp}
			\frac{k_\perp^5}{(k_\perp^2+k_z^2)^2k_z} P(\vec{k},z)\times 
			\nonumber \\ &[J_0(k_\perp r_p)+J_4(k_\perp r_p)] \label{eqn:xipp}
		\end{align}
		The $k_\perp^2/k^2$ terms in Eqs.~\eqref{eqn:xigp} and~\eqref{eqn:xipp} are equivalent to the Kaiser factor with 
		 $\beta=-1$:
		\begin{equation}
			\frac{k_\perp^2}{k^2}=1-\frac{k_z^2}{k^2}=1-\mu^2
		\end{equation}
		Mathematically, the projection effects in \xigp\ and \xipp\ introduce similar factors in the
        power spectrum as the Kaiser effect. 
		The Kaiser factor with positive $\beta$ leads to compression along the line of sight direction, 
		while a
		negative prefactor for $\mu^2$ leads to compression along $r_p$. The final shape 
		of the correlation 
		function is also determined by the Bessel functions which are different for different correlation functions. In the 
		absence of RSD and shape projection factors, $J_0$ leads to an isotropic correlation
        function, while $J_2$ and $J_4$ lead to a 
		peanut-shaped function in $(r_p,\Pi)$ space ($J_2,J_4\rightarrow0$ for $r_p\rightarrow 0$).
		Following the methodology used to study RSD \citep[see, e.g.,][]{Beutler2014}, we also compute 
		the monopole and quadrupole terms for the $\xi_{AB}$ using an expansion in Legendre 
		polynomials.
		\begin{align}\label{eqn:multipole}
		       \xi^l(r)&=\frac{2l+1}{2} \int \mathrm{d}\mu_r\, L^l(\mu_r)\xi(r,\mu_r)
	       \end{align}
	       $r=\sqrt{r_p^2+\Pi^2}=s$ (in redshift space) is the 3D separation of the galaxy pair, while 
	       $\mu_r=\Pi/r$ is the cosine of the angle between $\Pi$ and $r$. 
	       
	       In the data, we compute $\xi_{AB}$ as a function of $s$ and $\mu$, and apply the transform 
	       defined in Eq.~\eqref{eqn:multipole}. For theory 
	       calculations, we first compute $\xi_{AB}$ in the ($r_p,\Pi$) plane on a grid, then
           transform it to the ($r,
	       \mu$) plane and  apply the transform defined in Eq.~\eqref{eqn:multipole}. Since the theory 
	       prediction is calculated on a grid, there is 
	       some noise due to the finite grid size and Fourier ringing. Even though we smooth out the noise, 
	        we do not attempt to fit the theory to the observed multipoles. Instead, we use the best-fitting parameters 
	        from fits to $w_{AB}$, and simply compare the theory predictions with the data to
            confirm whether the observed trends in $(r_p,\Pi)$ are consistent with the above model
            on scales large enough that {effects from non-linear clustering and non-linear RSD are not important}.
	
\section{Data}\label{sec:data}
		 The SDSS \citep{2000AJ....120.1579Y} imaged roughly $\pi$ steradians
		of the sky, and the SDSS-I and II surveys followed up approximately one million of the detected
		objects spectroscopically \citep{2001AJ....122.2267E,
		  2002AJ....123.2945R,2002AJ....124.1810S}. The imaging was carried
		out by drift-scanning the sky in photometric conditions
		\citep{2001AJ....122.2129H, 2004AN....325..583I}, in five bands
		($ugriz$) \citep{1996AJ....111.1748F, 2002AJ....123.2121S} using a
		specially-designed wide-field camera
		\citep{1998AJ....116.3040G} on the SDSS Telescope \citep{Gunn2006}. These imaging 
		data were used to create
		the  catalogues of shear estimates that we use in this paper.  All of
		the data were processed by completely automated pipelines that detect
		and measure photometric properties of objects, and astrometrically
		calibrate the data \citep{Lupton2001,
		  2003AJ....125.1559P,2006AN....327..821T}. The SDSS-I/II imaging
		surveys were completed with a seventh data release
		\citep{2009ApJS..182..543A}, though this work will rely as well on an
		improved data reduction pipeline that was part of the eighth data
		release, from SDSS-III \citep{2011ApJS..193...29A}; and an improved
		photometric calibration \citep[`ubercalibration',][]{2008ApJ...674.1217P}.

		\subsection{Redshifts}
			Based on the photometric catalog,  galaxies are selected for spectroscopic 
			observation 
			\citep{Dawson:2013}, and the BOSS spectroscopic survey was performed
			\citep{Ahn:2012} using the BOSS spectrographs \citep{Smee:2013}. Targets
			are assigned to tiles of diameter $3^\circ$ using an adaptive tiling
			algorithm \citep{Blanton:2003}, and the data were processed by an
			automated spectral classification, redshift determination, and parameter
			measurement pipeline \citep{Bolton:2012}.

			We use SDSS-III BOSS data release 11 \citep[DR11;][]{SDSS2015}
			LOWZ galaxies, in the redshift range $0.16<z<0.36$. 
			The LOWZ sample consists of Luminous Red Galaxies (LRGs) at $z<0.4$, selected 
			from the SDSS DR8 imaging data and observed 
			spectroscopically in the BOSS survey. The sample is approximately volume-limited in the 
			redshift range $0.16<z<0.36$, with a number 
			density of $\bar{n}\sim 3\times10^{-4}~h^3\text{Mpc}^{-3}$ \citep{Manera2015}. We 
			combine the spectroscopic redshifts from BOSS with galaxy shape measurements 
			from \cite{Reyes2012}. BOSS DR11 has 225334 LOWZ galaxies within our redshift 
			range. However, \cite{Reyes2012} masks out certain regions that have higher Galactic 
			extinction, leaving us with 
			173855 galaxies for our LOWZ density 
			sample. 
	
	\subsection{Subsamples}
		To test for the dependence of our results on galaxy properties such as luminosity, color, and redshift, we 
		split the LOWZ sample into subsamples based on these properties following the 
		methodology detailed in \pI. To summarize, we define 
		four subsamples based on luminosity, $L_1$--$L_4$, with $L_1$ ($L_4$) being the brightest
        (faintest) subsample. Luminosity cuts are applied in 10 redshift bins, with $L_1$--$L_3$
        each containing 
		20\% of the galaxies and $L_4$ containing 40\% to improve the $S/N$ for the fainter 
		galaxies. Similarly, we define color subsamples $C_1$--$C_5$, each containing 
		20\% of the galaxies, with $C_1$ being the reddest subsample. For redshift, we 
		define two subsamples, $Z_1~(z\in[0.16,0.26])$ and $Z_2~(z\in[0.26,0.36])$.

		We also identify the galaxies in groups using the counts-in-cylinders (CiC) method \citep{Reid2009} and split the 
		sample into field galaxies (group of one), BGG (brightest group galaxy) and satellites (all
        non-field and non-BGGs). See \pI\ for more details and caveats related to the CiC group identification.
	
	\subsection{Shapes}\label{ssec:shapes}
		To measure the intrinsic alignments, we need to measure the shapes of the galaxies to be
        used in estimates of the ensemble intrinsic shear. One of the main goals of this 
		work is to compare different shape measurement methods and their impact on the 
		measured IA signal. Here we briefly describe the three different shape measurements used in this 
		paper. \referee{Though there are many shape measurement methods that are used for weak
          lensing in the literature 
		\citep[see for example][]{Mandelbaum2015}, our choice in this work is limited to the methods
        that (a) are 
		available for SDSS data and (b) have been previously used in intrinsic alignment studies in
        the SDSS.  However, we attempt to derive more general conclusions that could be applicable
        to other shape measurement methods by considering the
        essential properties of these methods (e.g., the choice of radii that they
      are sensitive to within the galaxy light profiles).}

		 Note that we do not have usable shape estimates from all methods for all the galaxies. In order to do a 
		fair comparison between different methods, the final shape
		sample in this work only contains the galaxies for which shapes are available from all three
        		methods. {Due to differences in the sky coverage of different shapes 
		(see Sec.~\ref{sssec:shapes_Isophotal}) we mask out some additional area on the sky, also reducing 
		the size of our density sample.} The final shape (density) sample has 122513 (131227) galaxies, making it 
		smaller and somewhat intrinsically brighter on average than the shape (density) sample used in \pI.
		
		\subsubsection{Re-Gaussianization Shapes}

		Re-Gaussianization shapes were used in the IA study done in \pI. These shape measurements 
		are described in more detail in \cite{Reyes2012}. Briefly, these shapes are measured 
		using the re-Gaussianization technique developed by \cite{Hirata2003}. The 
		algorithm is a modified version of ones that use ``adaptive moments'' (equivalent to fitting
        the light intensity profile to an elliptical Gaussian), determining shapes of the
        PSF-convolved galaxy image based on adaptive moments and then correcting the resulting
        shapes based on adaptive moments of the PSF.   The re-Gaussianization method involves
        additional steps to correct for  non-Gaussianity of both the PSF and the galaxy surface
        brightness profiles \citep{Hirata2003}. The components of the distortion are defined as
		\begin{equation}\label{eqn:distortion}
			(e_+,e_\times)=\frac{1-(b/a)^2}{1+(b/a)^2}(\cos 2\phi,\sin 2\phi),
		\end{equation}
		where $b/a$ is the minor-to-major axis ratio and $\phi$ is the position angle of the major 
		axis on the sky with respect to the RA-Dec coordinate system. The ensemble average of the
        distortion is related to the shear as
		\begin{align}
			\gamma_+,\gamma_\times&=\frac{\langle e_+,e_\times\rangle}{2\mathcal 
			R}\label{eqn:regauss_shear}\\
			\mathcal R&=1-\frac{1}{2}\langle e_{+,i}^2+e_{\times,i}^2-2\sigma_i^2\rangle\label{eq:R}
		\end{align}
where $\sigma_i$ is the per-component measurement uncertainty of the galaxy distortion, and 
		${\mathcal R}$ is the shear responsivity representing the response of an ensemble of
        galaxies with some intrinsic distribution of distortion values
		to a small shear \citep{Kaiser1995,Bernstein2002}. In \pI\ we used an ${\mathcal
          R}$ value for the entire SDSS shape sample 
		from \cite{Reyes2012}, $\mathcal R\sim0.87$, 
		whereas using  Eq.~\eqref{eq:R} we find $\mathcal R\sim0.925$ for the LOWZ sample, which we use in this 
		paper. In order to compare with \pI, the results in that paper should be rescaled by
        $0.87/0.925\approx 0.94$. 
		For different subsamples, we calculate $\mathcal R$ values separately using Eq.~\eqref{eq:R}, though the 
		variations between our subsamples are $\lesssim1\%$, well below the statistical errors.
		
		\subsubsection{Isophotal Shapes}\label{sssec:shapes_Isophotal}
			Isophotal shapes are measured in general by fitting an ellipse to an outer isophote of 
			the observed galaxy light profiles. SDSS DR7 provides isophotal shapes\footnote{http://classic.sdss.org/dr7/
			algorithms/classify.html} of galaxies using the
            isophote corresponding to a surface brightness of  
			25 mag$/$arcsec$^2$. These  have been used for several measurements of intrinsic
            alignments in the SDSS \citep{Okumura2009,Hao2011,Zhang2013}. The LRG intrinsic
            alignments measurements by \cite{Okumura2009} were later interpreted by \cite{Blazek2011} in 
			the context of the NLA model. \cite{Hao2011} measured satellite alignments in a
			sample of galaxy clusters. They found that the satellite alignment signal depended on
            the BCG apparent magnitude, 
			which they attributed to contamination from BCG light leaking into satellite isophotal shapes.
			In addition, isophotal shapes are not corrected for the PSF. 
			Due to the unreliability of these isophotal measurements, they were not included in the
            SDSS data release 8 (DR8) or subsequent data 
			releases. Thus, the isophotal shapes we use for this study do not cover the part of the LOWZ
            area coverage for which photometric measurements were made during DR8.
			
			We use the isophotal shape 
			parameters \texttt{isoA} (semi-major axis, $a$), \texttt{isoB} (semi-minor axis, $b$) and \texttt{isoPhi} (major axis 
			position angle, $\phi$) from the SDSS DR7 sky server. We get good isophotal shape 
			measurements for $\sim 297,000$ 
              galaxies ($\sim71\%$) in the full LOWZ sample, of which $122513$ are in the redshift range ($0.16<z<0.36$) 
              that we use for final analysis.  
			The ellipticity is defined as
			\begin{equation}\label{eqn:ellipticity}
				(\varepsilon_+,\varepsilon_\times) =\frac{1-b/a}{1+b/a}
				(\cos2\phi,\sin2\phi).
			\end{equation}
			The ensemble average of the ellipticity is an
            estimator for the shear:
			\begin{equation}\label{eqn:isoph_shear}
				\gamma_+,\gamma_\times={\langle \varepsilon_+,\varepsilon_
				\times\rangle}.
			\end{equation}
			
		\subsubsection{de Vaucouleurs Shapes}
			The \dev\ shapes are defined by fitting galaxy images to a \dev\ profile\footnote{https://
			www.sdss3.org/dr10/algorithms/magnitudes.php}
			\citep{Stoughton2002},
			\begin{equation}
				I(r)=I_0\exp\{-7.67[r/r_\text{eff}]^{1/4}\}
			\end{equation}
			with half-light radius $r_\text{eff}$. 
			The profile is allowed to have arbitrary axis ratio and position angle, and is truncated 
			beyond $7r_\text{eff}$ to go smoothly to zero beyond $8r_\text{eff}$.
			It is convolved with a double Gaussian approximation to the PSF model 
			\citep[for more details, see][]{Stoughton2002}. To reduce computation time, the fitting uses
			pre-computed tables of models \citep[][]{Lupton2001}, resulting in discretization in the
	            resulting model parameters \citep{Stoughton2002}.  
			The fit yields axis ratios and position angles that can 
			 be used to define ellipticities as in Eq.~\eqref{eqn:ellipticity}. 
			 
			 {
			 Several systematics in shape measurements depend on the apparent size of the galaxies.  The apparent size can quantified using the circularized radius of the galaxy, $R_\text{circ}$, 
			 defined as
	\begin{equation}\label{eqn:Rcirc}
		R_\text{circ}=R_\text{deVauc,r}\times\sqrt{\frac{b}{a}}
	\end{equation}
	We use the \dev\ profile fits in the  $r$ band to get the circularized
    radius. $R_\text{deVauc,r}$ is the \dev\ semi-major axis of the galaxy, and ${b/a}$ is the minor
    to major axis ratio for the \dev\ fit (already used to calculate the ellipticity). 
	}

		\cite{Li2013} used \dev\ shapes to measure the intrinsic alignment signal for the
        higher-redshift BOSS CMASS sample. One of their systematics tests consisted of randomly
        permuting the position angle measurements, yet after this permutation, they found
        that large scale  correlations of galaxy position angles persist. They attributed these
        correlations to the effects of cosmic variance and survey geometry, though these systematics
        can also arise from the incorrect PSF correction which can introduce such large scale
        correlations assuming that the PSF is roughly coherent across much of the survey.  The SDSS
        PSF has a preferred direction determined by the scan direction, and thus it is possible for
        coherent systematics to be present at some level even after randomly permuting the
        galaxies. We will explore systematic effects in the \dev\ shapes in detail in Sec.~\ref{sec:results}.

 		\subsubsection{PSF shapes}
		PSF models are defined at the position of stars \citep{Lupton2001,Stoughton2002}, then interpolated to
        arbitrary positions. For each frame, the PSF is expanded into Karhunen-Lo\'eve (K-L) basis using stars in the frame and its surrounding neighbours. The K-L models can then be 
		interpolated to the positions of galaxies to provide an image of the PSF at those
        locations.  We use \texttt{mE1psf\_r} and \texttt{mE2psf\_r} (defined using adaptive
        moments) from the SDSS
        sky server, and rotate them to the coordinate system
        defined by right ascension and declination. Since these quantities 
        are defined as in Eq.~\eqref{eqn:ellipticity}, we correct them  as in 
        Eq.~\eqref{eqn:regauss_shear} using $\mathcal R=1$ to get the PSF shear, $\gamma_{\text{PSF}}$.

		\subsubsection{Shape measurement systematics}\label{sssec:systematics_data}

			As described earlier, isophotal shapes are not corrected for PSF effects,
			 \dev\ shapes are approximately corrected using a double Gaussian PSF model, while
             re-Gaussianization uses the full PSF model.
The PSF causes an overall rounding of galaxy shapes, which if left uncorrected can introduce a
multiplicative bias $m$.  However, if the (coherent) PSF anisotropies leak into the galaxy shapes,
they produce an additive term $a\gamma_\text{PSF}$ that contributes 
spurious shape correlations. Thus, individual galaxy shear estimates 
$\gamma_\text{obs}$ include a purely random component (shape noise) along with contributions from
intrinsic shear $\gamma_I$ and systematics:
$$\gamma_\text{obs}=(1+m)\gamma_I+a\gamma_\text{PSF}.$$
These enter the ensemble intrinsic alignments observables as 
		\begin{align}
			&\langle g\gamma_\text{obs}\rangle=(1+m) \langle g\gamma_{I}\rangle+
			a\langle g\gamma_\text{PSF}\rangle=(1+m) \langle g\gamma_{I}\rangle\label{eqn:wgp_obs}\\
			&\langle \gamma_\text{obs}\gamma_\text{obs}\rangle= (1+m)^2 \langle\gamma_{I}
			\gamma_{I}\rangle+a^2\langle  \gamma_\text{PSF}\gamma_\text{PSF}\rangle.\label{eqn:wpp_obs}
		\end{align}
Eqns.~\eqref{eqn:wgp_obs} and~\eqref{eqn:wpp_obs} correspond to $\xi_{g+}$ and $\xi_{++}$,
respectively, and we have assumed that $\langle \gamma_I \gamma_\text{PSF}\rangle=\langle
g\gamma_\text{PSF}\rangle=0$.  That is, the intrinsic shear and PSF anisotropies are uncorrelated
(since they arise due to completely different physics), and likewise the PSF anisotropies are
uncorrelated with galaxy overdensities.  
		Incorrect PSF correction can clearly bias the IA measurements, with multiplicative bias mimicking 
			the behavior of IA amplitude $A_I$ and additive bias adding in a spurious term to
            {shape-shape correlations such as in $\xi_{++}$}. A simple way to detect the 
			effect of additive bias is to compute cross-correlations between the PSF and 
			galaxy shapes and normalize by the PSF shape auto-correlation function, which gives an approximately 
			scale-independent ratio. To summarize this in one number, we define
			\begin{equation}\label{eqn:Apsf}
			A_\text{PSF}=\frac{\int \mathrm{d}r_p \langle\gamma_\text{obs}\gamma_\text{PSF}\rangle/\langle
			\gamma_\text{PSF}\gamma_\text{PSF}\rangle}{\int \mathrm{d}r_p}\sim a 
			\end{equation}
			Multiplicative bias, on the other hand, is more difficult to detect using the data
            alone, 
            and requires detailed characterization for each shape measurement method using
            simulations, which is beyond the scope of this work.  
              
              Besides PSF-related systematics, several other effects such as incorrect sky subtraction and deblending can also 
              lead to spurious alignment signals. Some of 
              these systematics lead to biases similar to those already discussed, 
              while {many} others could be revealed by systematic tests that we present in 
              Sec.~\ref{ssec:systematics}.

		\subsubsection{Physical effects}\label{sssec:physical_data}
Aside from systematics, there are physical reasons why different shape measurements can give
different results.  
			Several studies of bright elliptical galaxies have found evidence 
			for ellipticity gradients and isophote twisting with radius \citep[see, e.g.,][]
{Wyatt1953,Tullio1978,Tullio1979,Kormendy1982,Fasano1989,Nieto1992,Romanowsky1998,Lauer2005,Pasquali2006}. 
			\cite{Tullio1978,Tullio1979} first found environment-dependence of ellipticity
            gradients. Galaxies for which the ellipticity increases (decreases) with radius 
			 are found preferentially in dense (isolated) environments. 
			\cite{Pasquali2006} confirmed these trends in a sample of 18 elliptical galaxies from
            the {\em Hubble Space Telescope} (HST) Ultra-deep Field (UDF).  
			\cite{Wyatt1953} also found evidence of isophote twisting in elliptical galaxies 
			\citep[see also][]{Fasano1989,Nieto1992,Lauer2005}.
			These observations are consistent with bright ellipticals being triaxial systems with varying 
			triaxiality 
			as a function of radius, which in projection appears as isophote twisting \citep[e.g.,][]
			{Romanowsky1998}. However, external tidal fields can also 
			influence the galaxy shapes leading to isophote twisting  \citep{Kormendy1982}, which
            could mean that the measured intrinsic alignments depend on the effective radius within
            the galaxy used for measuring the shape. IA measured with shapes that are weighted
            towards the galaxy centers (outer regions) will be weaker (stronger).
		
			Several studies have also looked at variations in galaxy/halo ellipticity 
			and IA response with the radius using simulations
           \citep{Schneider2013,Tenneti2014,Velliscig2015,Velliscig2015b}. The 
			consensus has been that halo or galaxy shapes get rounder with increasing radius
            \citep[for a contradictory result, see][]{Velliscig2015}  but the 
			outer regions also show stronger IA, primarily because of isophotal twisting.

			Different radial weighting in the various shape measurements can in principle allow 
			us to test these variations from data, although interpretation of our results will be  
			complicated by the possibility of systematics such as PSF contamination as discussed earlier. 
			 The re-Gaussianization shapes assign 
			higher weights to inner regions of galaxy profiles, while \dev\
			profiles have a comparatively somewhat broader coverage in radius, and isophotal shapes only measure the outer 
			regions of the galaxy light profiles. If tidal fields lead to isophote twisting, making outer 
			regions of the galaxy more intrinsically aligned, we should expect higher IA signal for isophotal shapes 
			followed by \dev\ and re-Gaussianization shapes.

\section{Results}\label{sec:results}

To begin, we explore the features of the galaxy shape distributions (Sec.~\ref{ssec:pe}) and carry
out basic systematics tests of intrinsic alignments two-point correlation functions
(Sec.~\ref{ssec:systematics}) using the different shape measurements.  The results in these
subsections provide some basic context that will be useful when interpreting the intrinsic
alignments results in later subsections. Then we confirm the equivalence of intrinsic alignment
two-point statistics using distortions (Eq.~\ref{eqn:distortion}) vs.\ ellipticities
(Eq.~\ref{eqn:ellipticity}) in Sec.~\ref{ssec:edef}, and compare intrinsic alignments for different
shape measurement methods and IA estimators (Sec.~\ref{ssec:results_shapes}).  Our tests of the
redshift-space $(r_p,\Pi)$ structure of the IA two-point correlations are in
Sec.~\ref{ssec:results_anisotropy}.  Finally, we compare with other studies in
Sec.~\ref{ssec:comparison}.

   \subsection{Ellipticity from different shape measurement methods}\label{ssec:pe}

\begin{figure*}
	      	\begin{subfigure}{\columnwidth}
		         \centering
		         \includegraphics[width=.9\columnwidth]{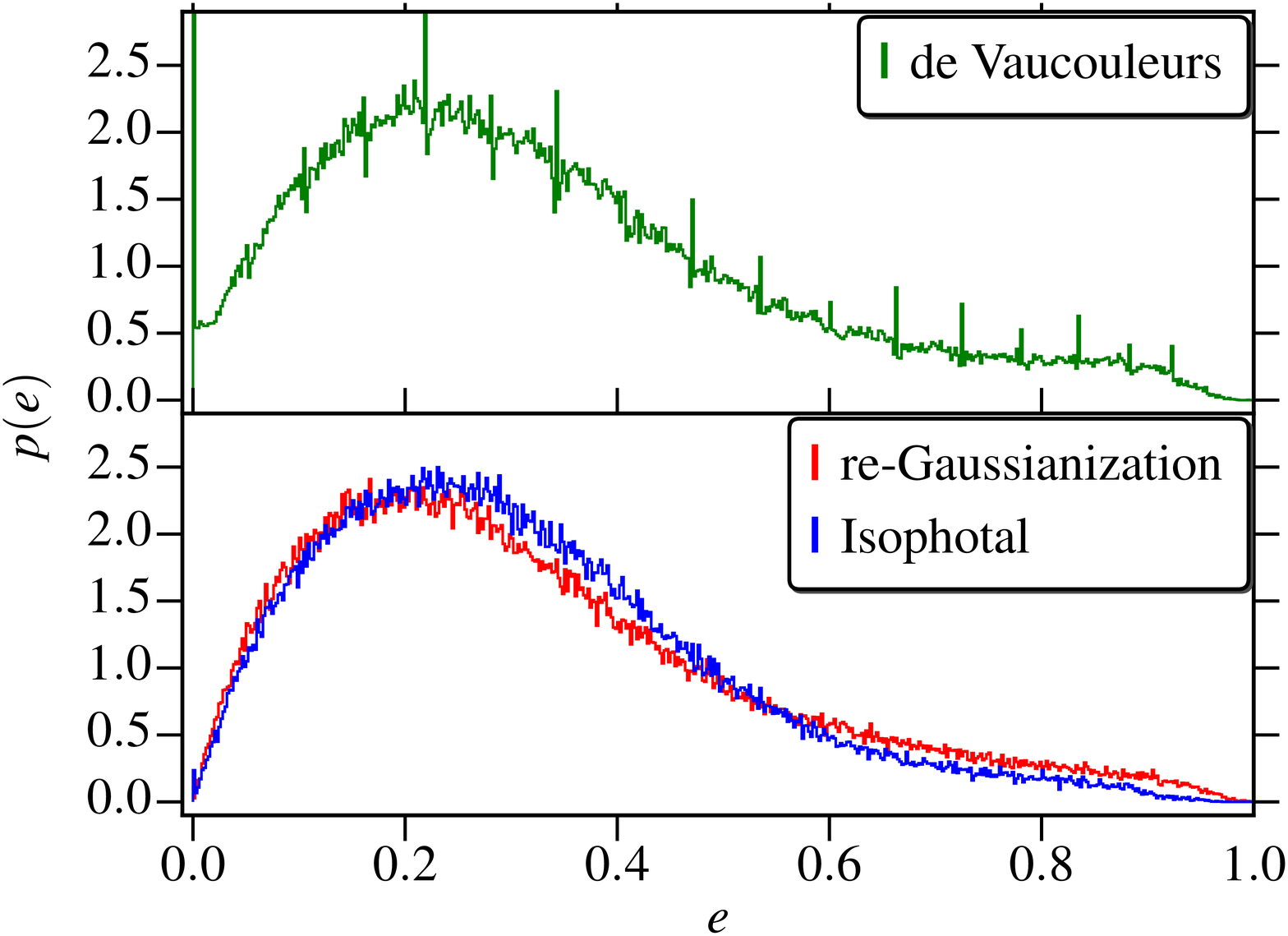}
		         \caption{}
		         \label{fig:e_hist}
	         \end{subfigure}
	      	\begin{subfigure}{\columnwidth}
		         \centering
		         \includegraphics[width=.9\columnwidth]{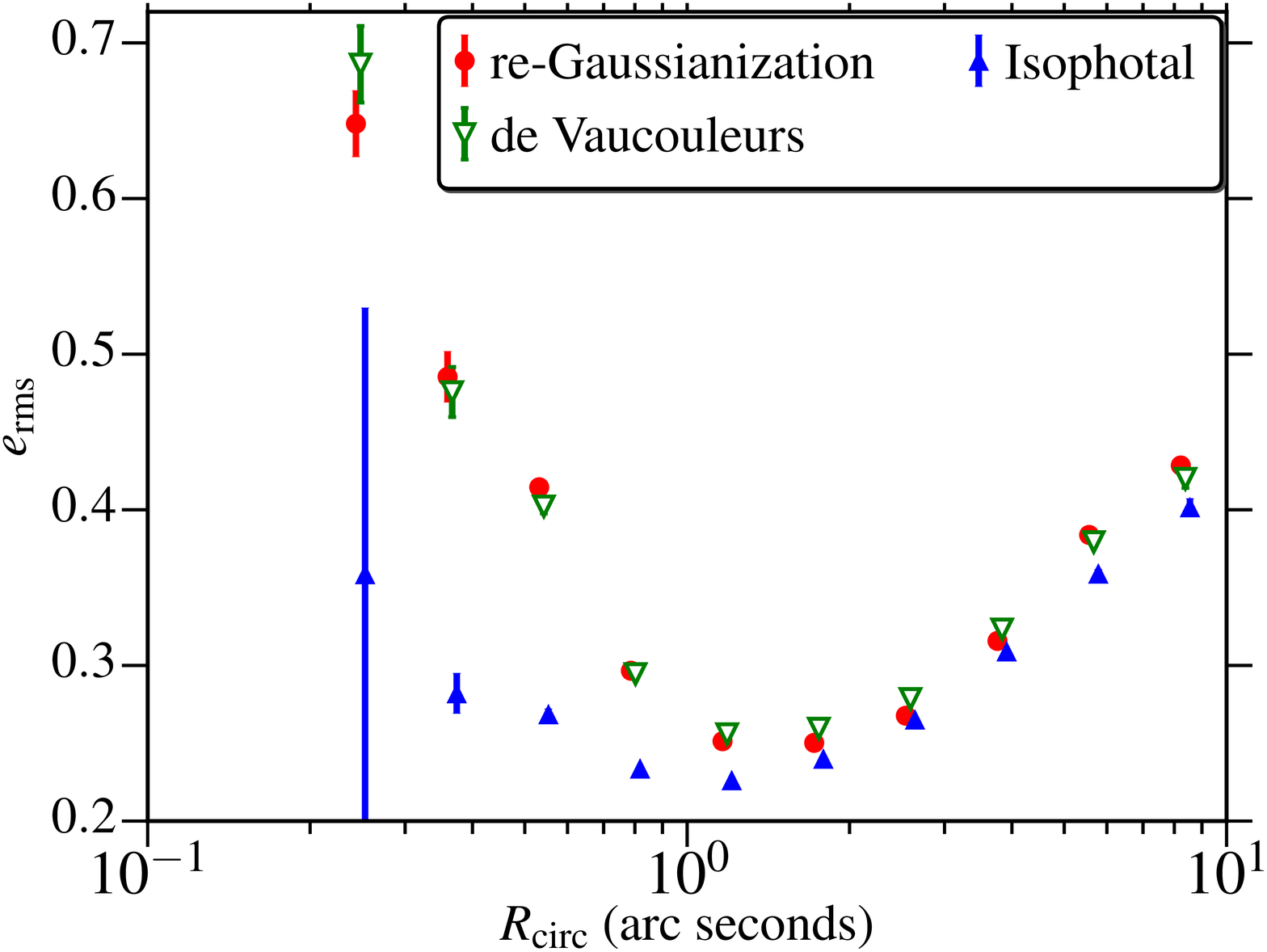}
		         \caption{}
		         \label{fig:erms_Rcirc}
	         \end{subfigure}
		 \caption{(a) Probability distribution of $e$ for different shape measurement methods. The
           periodic spikes in the \dev\ $p(e)$ are \referee{discussed in the text of Sec.~\ref{ssec:pe}}. (b) The RMS ellipticity per shape component, $
		 e_{\text{rms}}$, for different shape measurement methods as a function of \referee{circularized apparent 
		 radius $R_\text{circ}$ (Eq.~\ref{eqn:Rcirc})}.
         	}		
		 \label{fig:erms}
	      \end{figure*}

   	In this section we study the distortion, $e$, as defined in Eq.~\eqref{eqn:distortion}, using
    	re-Gaussianization, isophotal, and \dev\ shapes. 
   	Fig.~\ref{fig:e_hist} shows the probability distribution $p(e)$, and Fig.~\ref{fig:erms_Rcirc} shows the 
	{per-component RMS distortion, \erms, as a function of 
	circularized apparent radius $R_\text{circ}$ for  each 
	method.
	}  The $p(e)$ for the \dev\ shapes in
    Fig.~\ref{fig:e_hist} has periodic spikes that are likely 
	caused by discretization of the model parameters during the fitting procedure
    \citep{Lupton2001,Stoughton2002}.  Similar quantization features are seen in the position angle
    distribution (not shown). 
	The \dev\ shapes have the highest RMS 
	distortion {($\erms=0.261$), followed closely by re-Gaussianization ($\erms=0.256$), with isophotal shapes ($\erms=0.241$) giving a noticeably
    lower RMS distortion}. Since the three methods measure galaxy shapes at different 
	effective radii, the straightforward interpretation of this trend is that the galaxy ellipticity
    first increases and then decreases with radius. However, the quantization in the \dev\ shapes
    combined with the fact that (as shown in 
	Sec.~\ref{ssec:systematics}) the \dev\ shapes have significant additive PSF bias casts doubt on
    this interpretation.  It may, however, be a valid interpretation of the differences between
    re-Gaussianization and isophotal shapes.

	The trend for re-Gaussianization vs.\ isophotal shapes is qualitatively in 
	agreement with hydrodynamic simulations 
	\citep{Tenneti2015b}, where the axis ratio increases with radius. 

The lower values of $e_\text{rms}$ in isophotal shapes is primarily driven by the deficit of galaxies with 
	$e>0.7$, as shown in Fig.~\ref{fig:e_hist}. This could also be due to some systematic effect such as multiplicative 
	bias from the PSF, 
{which tends to make galaxy shapes rounder and will be more important for more
 	elongated galaxies. The trends in $e_\text{rms}$  in  
	Fig.~\ref{fig:erms_Rcirc} are consistent with the presence of some systematic bias. The PSF effects become more 
	important for smaller galaxies (smaller $R_\text{circ}$), thus making them appear rounder and
    thereby reducing the incidence of high-ellipticity objects. The fact that the isophotal shapes,
    which have no PSF correction at all, have a lower \erms\ and that this difference between
    isophotal and other shapes gets progressively
    more pronounced for galaxies with a small apparent size is a clear signature of PSF dilution
    systematics. Note that (for a fixed total flux), the effect of pixel noise increases measurement
    errors preferentially for the smaller sizes,  leading to an increase in the \erms\ for smaller
    galaxies for all measurement methods.  While we have attempted to subtract the measurement
    errors, these are known to be underestimated \citep{Reyes2012} and thus incompletely removed. 
	Fully disentangling true physical effects like ellipticity gradients from systematic effects requires detailed analysis of 
	different shape 
	measurements using simulations, which is beyond the scope of this work. Our statements
	about radial evolution in ellipticity using isophotal shapes are only valid under the assumption that they are not 
	significantly biased by the PSF or other systematics. In the context of the discussion so far, this is a reasonable 
	assumption since $\sim90\%$ of galaxies in our sample have $R_\text{circ}>1\arcsec$.}

		      \begin{figure*}
	      	\begin{subfigure}{\columnwidth}
	         \centering
	         \includegraphics[width=.9\columnwidth]{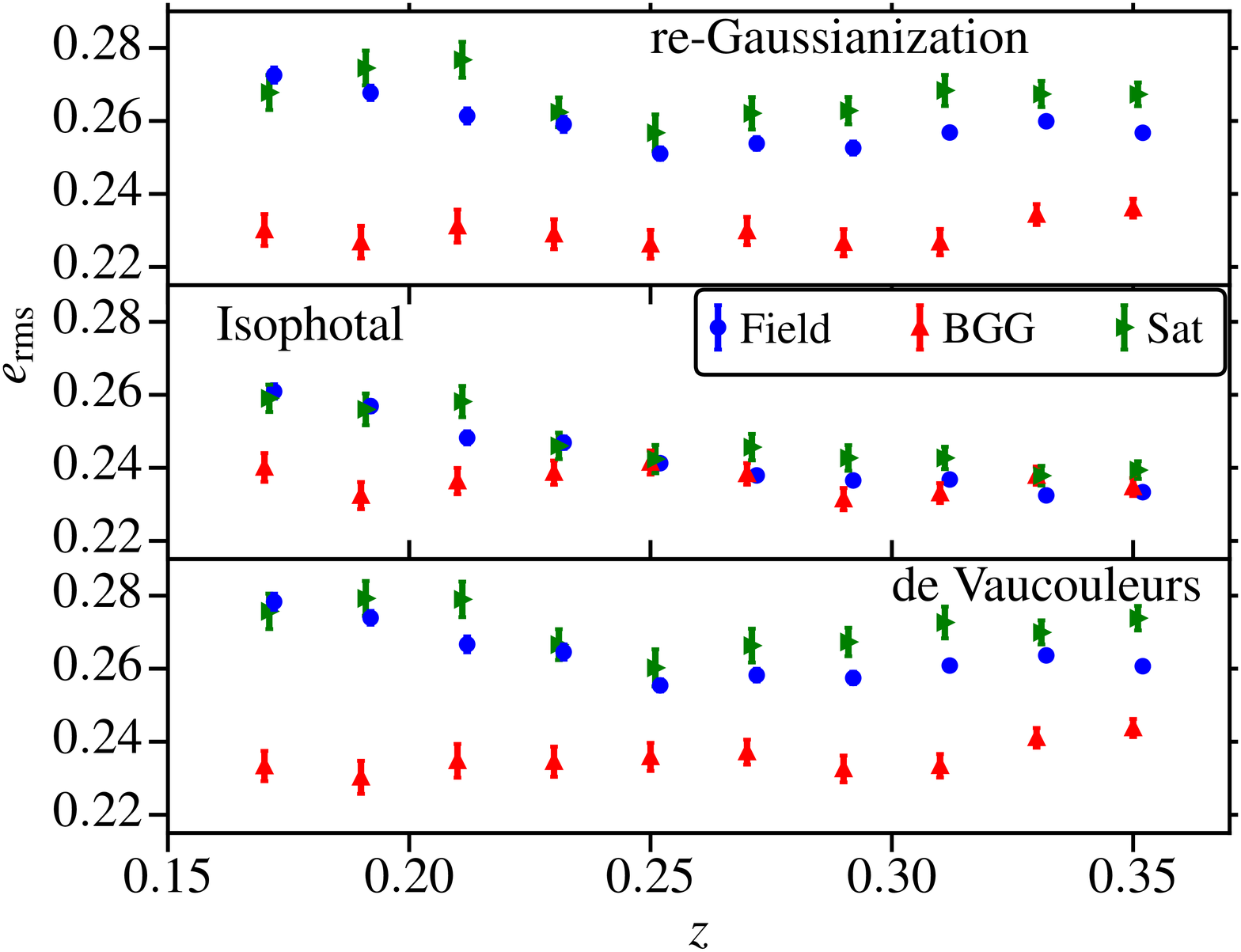}
	         \caption{}
	         \label{fig:erms_env}
	         \end{subfigure}
	         \begin{subfigure}{\columnwidth}
	          \centering
	         \includegraphics[width=.9\columnwidth]{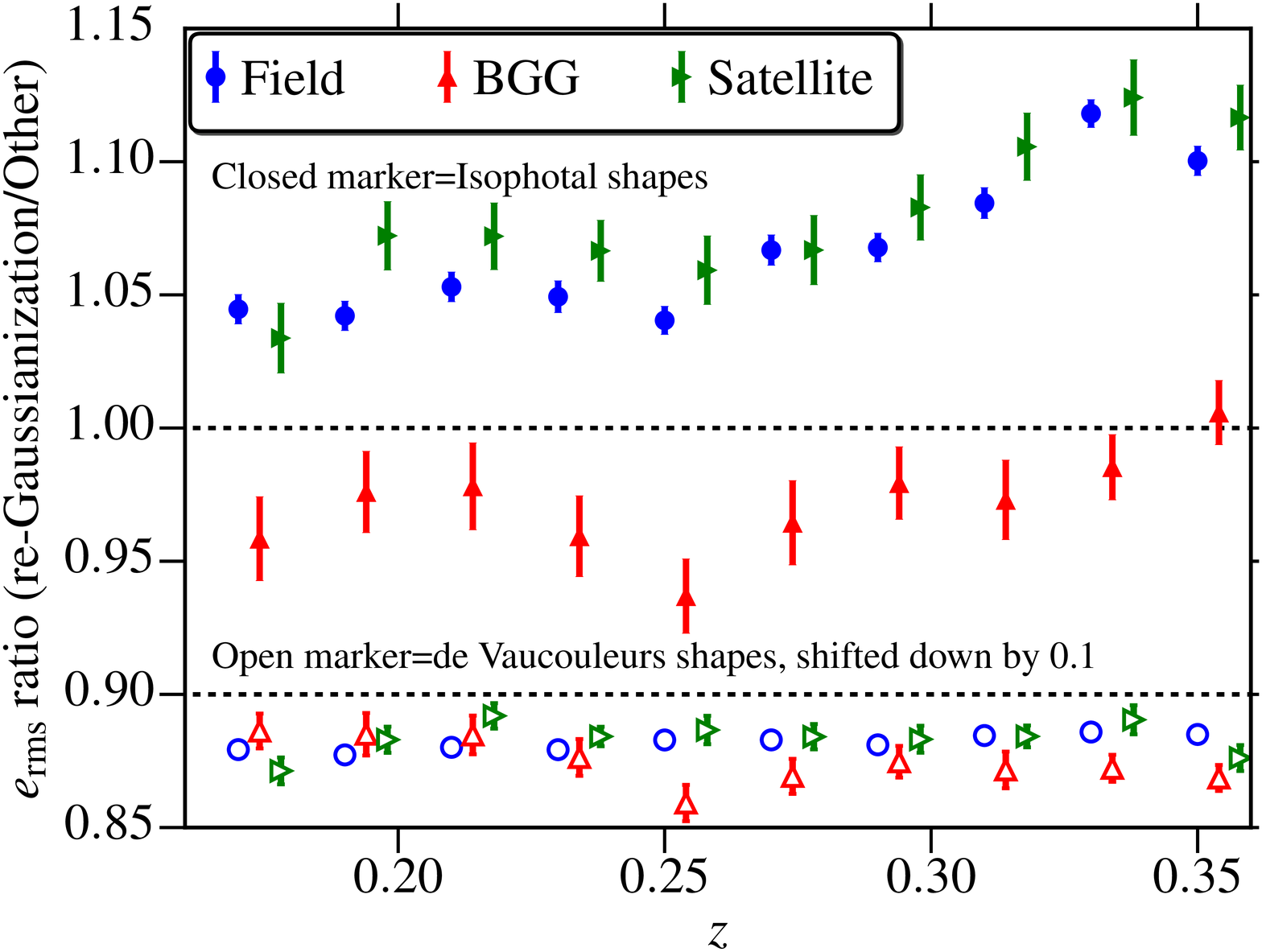}
	         \caption{}
	         \label{fig:erms_env_ratio}
		\end{subfigure}
		\caption{(a) The $e_{\text{rms}}$ as a function of redshift for galaxies in different environments using 
			different shape 
			measurement methods. 
			BGGs are rounder than satellite and field galaxies using all three methods, though the 
			differences are much less significant using isophotal shapes. 
			(b) Ratio of $e_{\text{rms}}$ using re-Gaussianization shape to isophotal or \dev\ shapes for 
			galaxies in 
			different environments. BGGs are relatively more elliptical in isophotal shapes while satellites and 
			field 
			galaxies are more round using isophotal shapes. The differences are dominated by ellipticity 
			differences
			using re-Gaussianization shapes as shown in (a). The variations in ellipticity between re-
			Gaussianization 
			and \dev\ shapes are nearly identical across all galaxy environments (open markers in (b)).
			}
		\label{fig:erms_env2}
	      \end{figure*}

	As described in Sec.~\ref{sssec:physical_data}, few studies have observed the ellipticity trends in small samples of 
	elliptical galaxies 
	\citep[see for eg.][]{Tullio1978,Tullio1979,Pasquali2006}. \cite{Tullio1978,Tullio1979} first identified environment 
	dependence of ellipticity gradients. Galaxies that have increasing ellipticity with radius are found preferentially in dense 
	environments like clusters and groups, while those that have decreasing ellipticity with radius were generally isolated though 
	they were also found in groups. 
	We study these trends in the much larger LOWZ sample by identifying BGGs, satellites and field
    galaxies. Fig.~\ref{fig:erms_env} 
shows the $e_{\text{rms}}$ for these different environment subsamples for all three shape measurements, 
	and   Fig.~\ref{fig:erms_env_ratio} shows the ratio of $e_{\text{rms}}$ from re-Gaussianization shapes 
	to isophotal and \dev\ shapes as a function of redshift. For both re-Gaussianization and \dev\
    shapes, BGGs are rounder than field and satellites galaxies, while the trend is less clear for isophotal shapes. 
	Doing the comparison between different shapes as shown in Fig.~\ref{fig:erms_env_ratio}, BGGs get
    more elliptical with increasing radius  
	while satellites and field galaxies get rounder, consistent with the observations of \cite{Tullio1979}. 
	
   \subsection{Systematics in two-point functions}\label{ssec:systematics}

\begin{figure}
         \centering
         \includegraphics[width=\columnwidth]{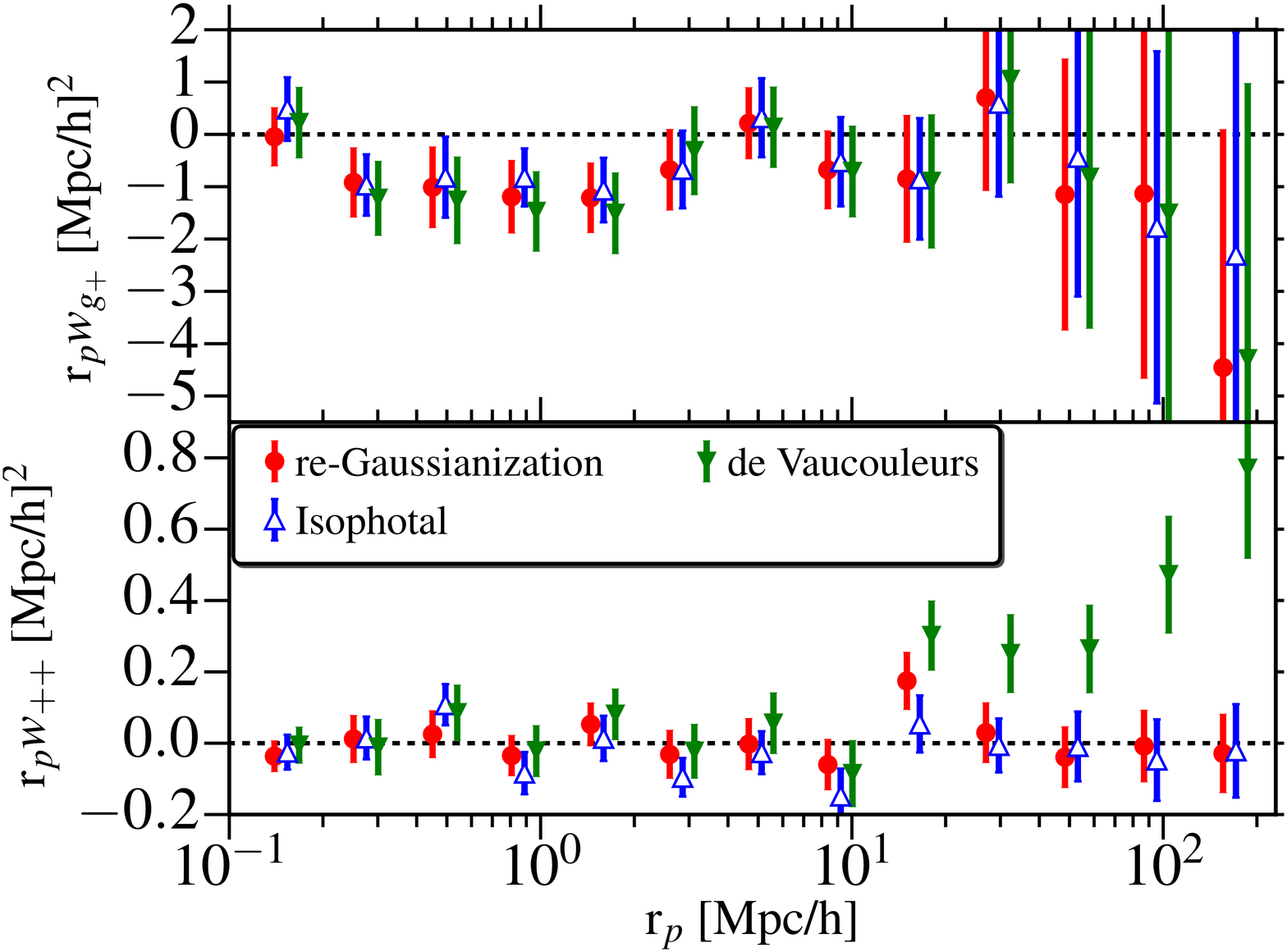}
         \caption{Intrinsic alignment correlation functions using integration only over large
  line-of-sight separations ($|\Pi|\in[200,500]$\mpch) for the full LOWZ sample, with different
  shape measurement methods.  The top and bottom show $r_p w_{g+}$ and $r_p w_{++}$, respectively.
         }
         \label{fig:wgp_wpp_large_pi}
      	\end{figure}

   	In this section we show some systematics tests in the intrinsic alignment two-point correlation
    functions. Fig.~\ref{fig:wgp_wpp_large_pi} shows the density-shape (\wgp) and shape-shape (\wpp)
    correlations using large $
	\Pi$ separations ($|\Pi|\in [200,500]$ \mpch). For such large separations, we do not expect any 
	contribution from intrinsic alignments, so a deviation of the signals from zero is more likely
    from additive systematics due to the PSF (in the case of \wpp) or, on small scales, incorrect sky subtraction. These signals are mostly consistent with zero 
	(see Table~\ref{tab:large_pi_significance} for $\chi^2$ and $p$ values) and thus do not indicate the presence of 
	systematics, except for \wpp\ with \dev\ shapes on large scales.  The slight negative signals in \wgp\ are
    consistent in magnitude and $r_p$-scaling with the presence of a small lensing signal (tangential shape alignments) given the
    redshift separations between the galaxy pairs and the known lensing signals from \pI.

		\begin{table}
                                \begin{tabular}{|c|c|c|c|}
                                \hline
                                Correlation&Shape & $\chi^2$& $p-$value \\ \hline
                                \wgp\ & re-Gaussianization&17.9&0.29\\
                                \wgp\ & Isophotal&17.4&0.31\\
                                \wgp\ &\dev\ &20.4&0.19\\
				\wpp\ &re-Gaussianization&9.8&0.79\\
                                \wpp\ &Isophotal&15.2&0.43\\
                                \wpp\ &\dev\ &31.6&0.02\\ \hline
				\end{tabular}
                                \caption{ $\chi^2$ and $p-$values for \wgp\ and \wpp\ calculated using large $\Pi$ separations 
                                ($|\Pi|\in [200,500]$ \mpch) and different shape measurements. All
                                signal are consistent with a null 
                                detection except for \wpp\ with \dev\ shapes.
                                }
                                \label{tab:large_pi_significance}
                        \end{table}

\begin{figure}
         \centering
         \includegraphics[width=\columnwidth]{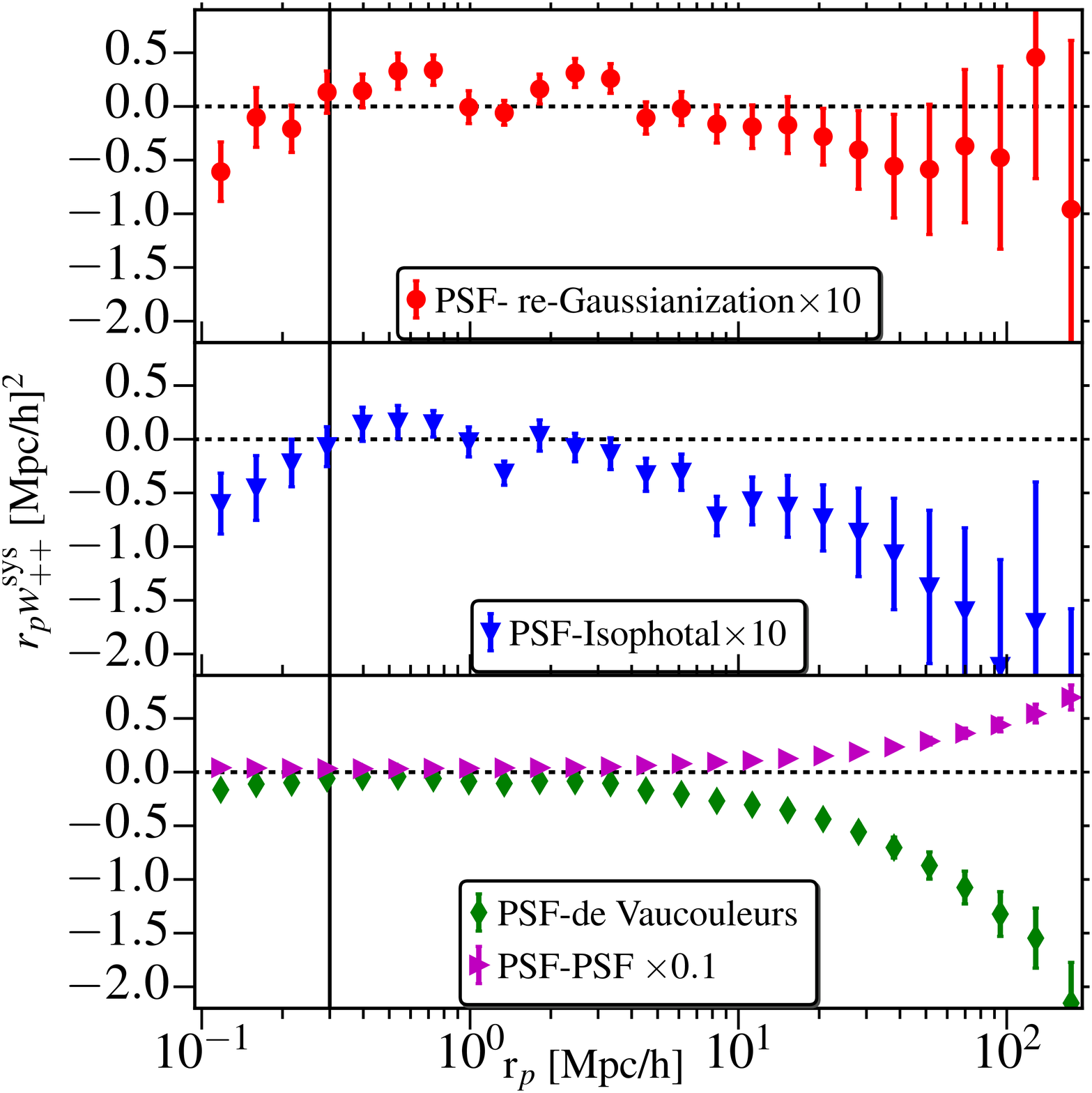}
         \caption{Galaxy shape vs.\ PSF shape cross-correlations using the full LOWZ sample
           with 
         different shape measurement methods, as a way to reveal additive PSF contamination in the
         galaxy shapes. The signal for re-Gaussianization and isophotal shapes has been multiplied 
         by a factor of 10 in this plot, while the signal for PSF-PSF correlations is suppressed by
         a factor of 10. The black solid line marks the SDSS fiber collision limit.
      }
         \label{fig:psf_wpp}
      \end{figure}

\begin{figure}
         \centering
         \includegraphics[width=\columnwidth]{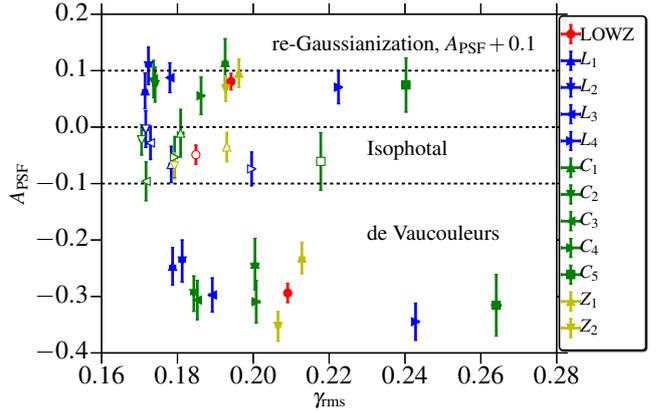}
         \caption{Galaxy shape vs.\ PSF shape cross-correlation amplitude, $A_\text{PSF}$
           (Eq.~\ref{eqn:Apsf}), using isophotal (open 
	         markers), re-Gaussianization (filled markers, $\apsf+0.1>0$, shifted up for clarity)
             and \dev\  
	         (filled markers, $\apsf<-0.1$) shapes. re-Gaussianization and isophotal shapes show relatively low and similar 
	         amplitudes while \dev\ shapes have a large value of \apsf, indicative of significant additive bias 
	         from the PSF anisotropy.}
         \label{fig:psf_amp}
      \end{figure}

	As discussed in Sec.~\ref{sssec:systematics_data}, additive PSF contamination in galaxy shape measurements can affect IA 
	results, particularly \wpp, if not corrected properly. To better understand this contamination, we directly calculate $\langle 
	\gamma_\text{obs}\gamma_\text{PSF}\rangle$ cross-correlations using PSF shapes at the 
	positions of galaxies. 
		Fig.~\ref{fig:psf_wpp} shows the cross-correlations of different shape measurements with the
        PSF shape, $w_{++}^\text{sys}$, with \dev\ 
		shapes showing strong correlations. These are likely the cause of the non-zero \wpp\ for
        large $\Pi$ in Fig.~\ref{fig:wgp_wpp_large_pi}.  Isophotal and re-Gaussianization shapes, on the other hand,
		 show much lower levels of contamination, which suggests that these two 
		 are not strongly affected by additive PSF errors. Fig.~\ref{fig:psf_amp} shows 
		 $A_\text{PSF}$ defined in Eq.~\eqref{eqn:Apsf} for different galaxy subsamples and shape measurement methods. This figure 
		 confirms our conclusion that \dev\ shapes have stronger additive PSF bias while isophotal 
		 and re-Gaussianization shapes have much smaller levels of
        		 contamination. Using the full LOWZ sample, we find $A_\text{PSF}=-0.019\pm0.015$, 
		 $-0.049\pm0.017$, and $-0.29\pm0.02$ for re-Gaussianization, isophotal, and \dev\ shapes.
		 Our value of $|A_\text{PSF}|$ for re-Gaussianization is smaller than that in \cite{Mandelbaum2015}, who found $A_\text{PSF}
		 \sim-0.1$; however, those results were for a simulated galaxy sample extending to much lower $S/N$ and resolution, which is expected to have 
		 relatively stronger additive systematics compared to LOWZ.

{It is quite interesting and somewhat counter-intuitive that the \dev\ shapes, which include PSF correction
  with an approximate PSF model,
  have a much larger $A_\text{PSF}$ than the isophotal shapes, which do not.  We propose that the
  reason for this is that the \dev\ shapes are weighted towards the central part of the galaxy light
profile and thus rely on small scales (which are highly sensitive to the PSF).  Thus, a small error
in the PSF model can be very important.  In contrast, the isophotal shapes use such a low
surface-brightness isophote that they correspond to much larger scales, where the impact of the PSF
is much smaller, and even without correction, $A_\text{PSF}$ can be quite small.}

		 There is no strong dependence of the additive PSF contamination on galaxy properties such as 
		 luminosity and color. The contamination in \dev\ shapes does show significant redshift dependence, 
		 with the higher redshift sample, Z2, showing stronger contamination; a similar but much less significant effect is 
		 present in isophotal and re-Gaussianization shapes. This is likely due the fact that higher 
		 redshift galaxies have a smaller apparent size, making the contamination from 
		  PSF anisotropy more important for such galaxies. There are hints of luminosity 
		 dependence to the additive PSF contamination for both \dev\ and isophotal shapes, though
         these trends are not very significant. $|A_\text{PSF}|$ is 
		 expected to be larger for lower luminosity samples since those galaxies generally have smaller 
		 apparent size, making PSF contamination more important.

		 We caution that the tests in this section do not rule out multiplicative bias, which could
         change the IA two-point correlation functions in a way that is degenerate with the IA amplitude $A_I$. 
	      
   
   \subsection{Ellipticity definition}\label{ssec:edef}

\begin{figure}
         \centering
         \includegraphics[width=\columnwidth]{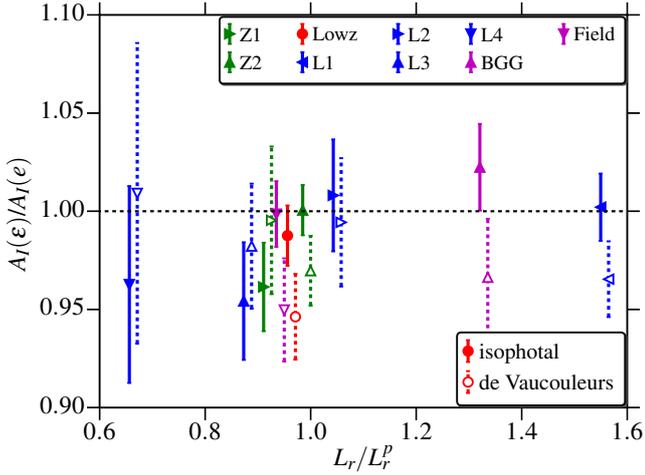}
         \caption{ Ratio of IA amplitude, $A_I$, obtained via NLA model fits to \wgp\ calculated using shears from two different ellipticity
           definitions, $e$ (Eq.~\ref{eqn:distortion}) and
           $\varepsilon$ (Eq.~\ref{eqn:ellipticity}), for isophotal (solid lines) and \dev\
           (dashed lines) shapes. Both definitions give consistent results across all luminosities,
           redshifts, and environments.
           }
         \label{fig:e_def_ratio}
      	\end{figure}

	To confirm the consistency of intrinsic alignments results using the two different ellipticity
    definitions in Eq.~\eqref{eqn:distortion} and~\eqref{eqn:ellipticity}, we calculate the
    density-shape correlation function \wgp\ with isophotal and 
	\dev\ shapes using both definitions, modifying the \wgp\ estimator appropriately. 
Fig.~\ref{fig:e_def_ratio} shows the ratio of the inferred $A_I$ for different subsamples from the
NLA model fits to \wgp\ using the 
	two shear estimators.  $A_I$ measured 
	using both definitions are consistent within 1$\sigma$ for isophotal shapes, and there 
	are 
	no strong dependences on galaxy properties such as color or luminosity. In principle, this is
    precisely as expected; however, differences could arise due to systematic errors in the RMS
    ellipticity calculation if the error estimates are incorrect.  For the 
	rest of the paper, we use $\varepsilon$ {(Eq.~\ref{eqn:ellipticity})} as our galaxy shape estimator for isophotal and \dev\ shapes.

   \subsection{IA with different shape measurements}\label{ssec:results_shapes}

   	In this section we present the IA measurements using the different shape measurement methods 
	discussed in Sec.~\ref{ssec:shapes}.

\begin{figure*}
	      \begin{subfigure}{\columnwidth}
		         \centering
	        		 \includegraphics[width=0.9\columnwidth]{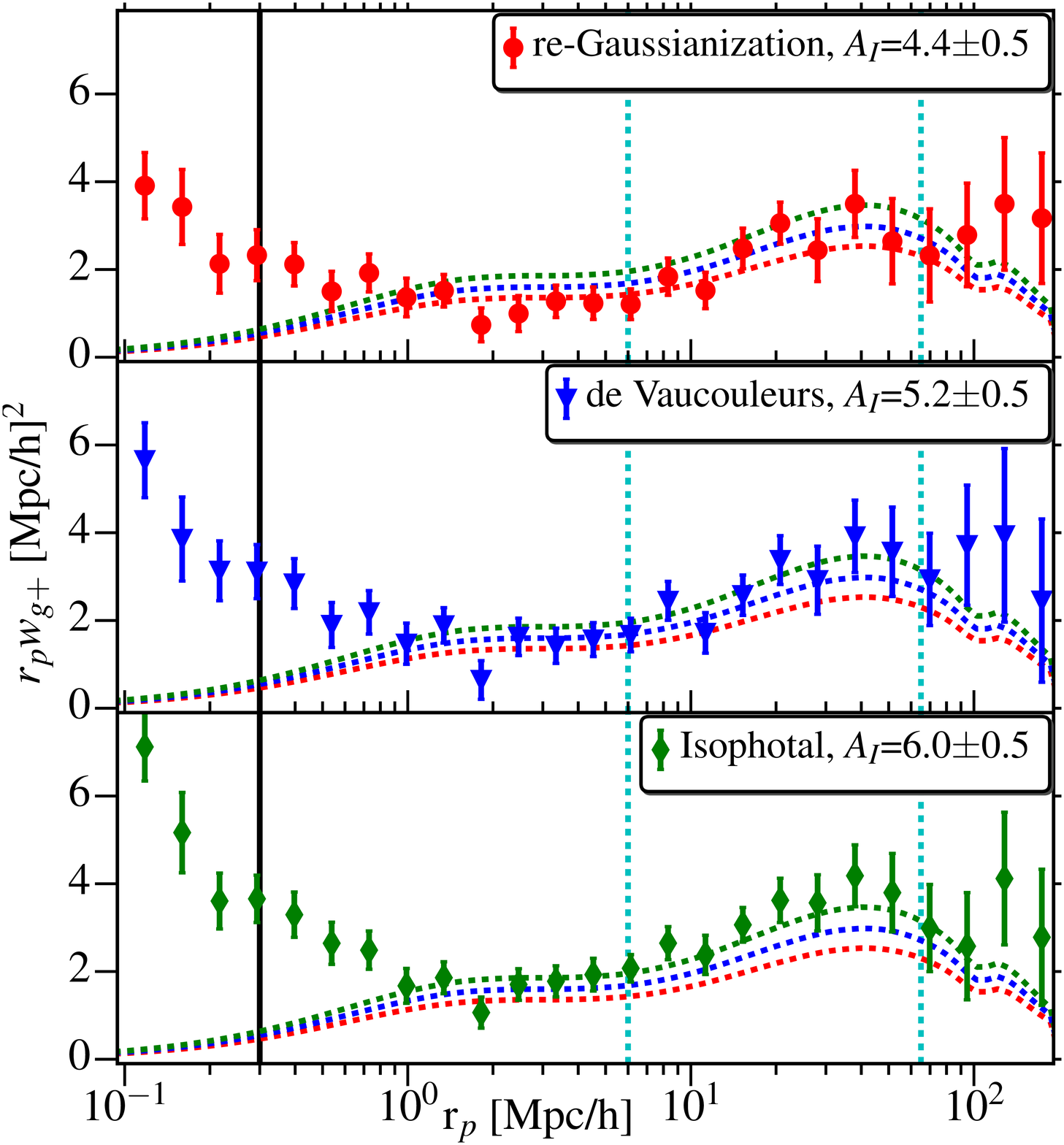}
		        	 \caption{}
		         \label{fig:lowz_wgp}
		\end{subfigure}         
		\begin{subfigure}{\columnwidth}
		         \centering
		         \includegraphics[width=0.9\columnwidth]{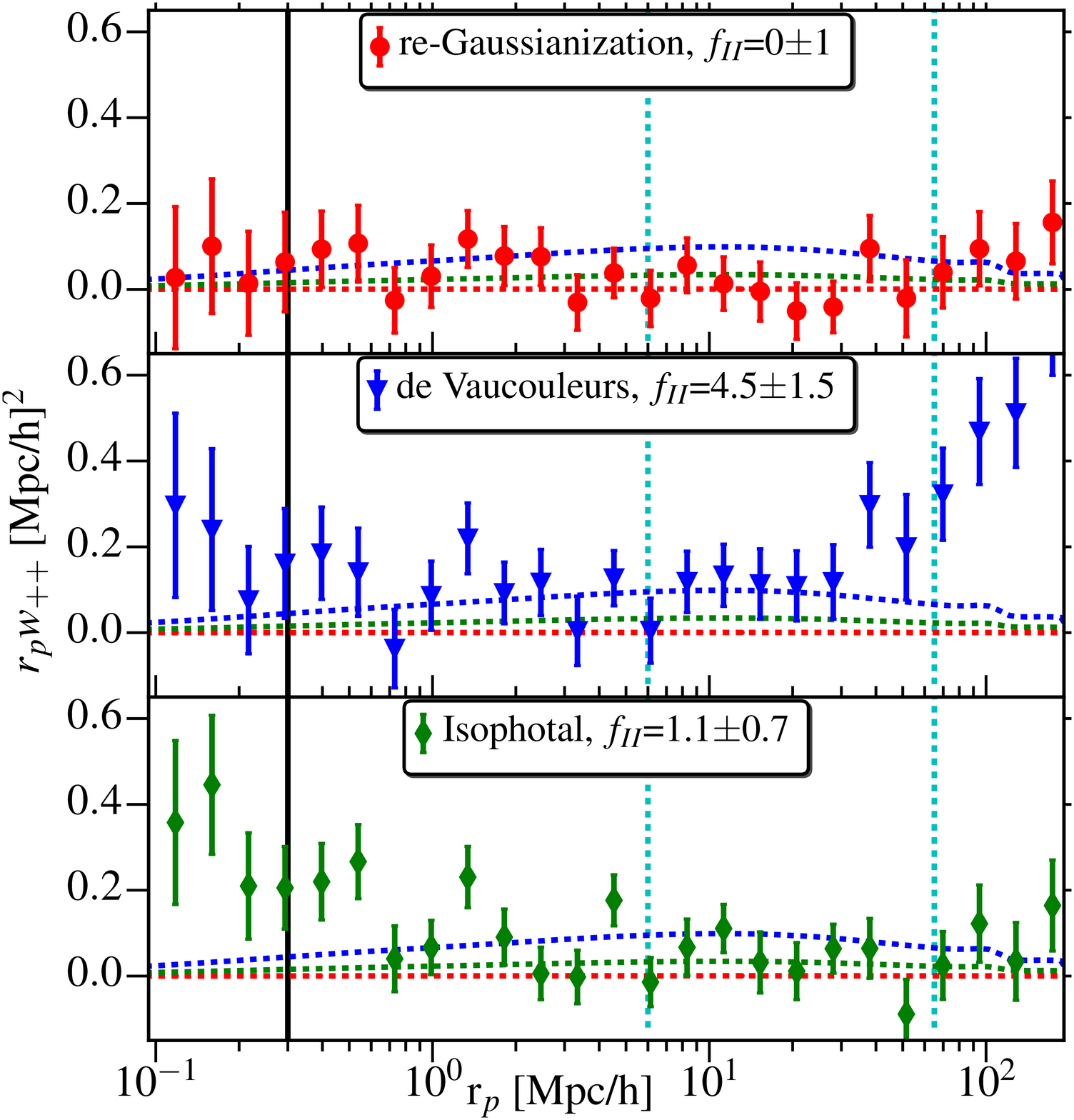}
			\caption{}
		         \label{fig:lowz_wpp}
	      \end{subfigure}
		         \caption{
		         (a) \referee{The projected galaxy density-shape correlation function \wgp\ (Eq.~\ref{eqn:wgp})} using the full LOWZ sample and different shape 
		         measurement methods,
	         	isophotal (blue), re-Gaussianization (red) and \dev\ shapes (green), along with the
       		         best-fitting  
			NLA models. For comparison, the 
			best-fitting NLA models for all three shape measurement methods are plotted on all panels. 
			Isophotal 
			shapes show the highest IA signal, followed by \dev\ and re-Gaussianization shapes (note that 
			error bars are correlated between the different shape measurement methods).
			The solid black line shows the SDSS 
			fiber collision limit, and the dashed cyan lines show the range of $r_p$ used for the NLA model 
			fitting. 
		         (b) Same as (a), but for \referee{the projected shape-shape correlation function \wpp\ (Eq.~\ref{eqn:wpp})}.
}
	      \end{figure*}

\begin{figure}
         \centering
         \includegraphics[width=\columnwidth]{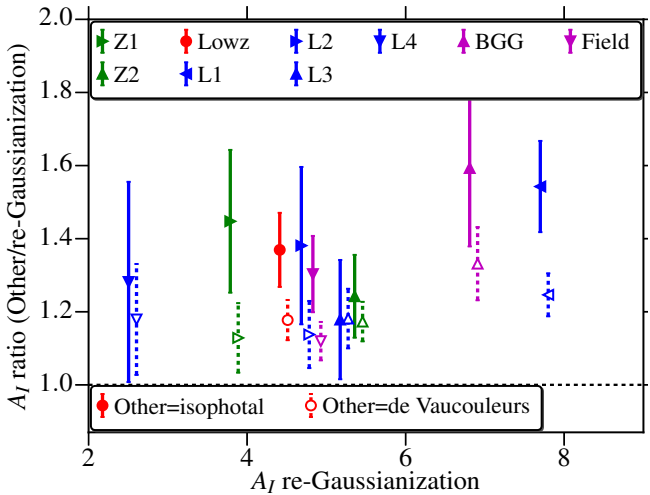}
         \caption{Comparison of NLA model amplitude $A_I$ for isophotal and \dev\ 
         vs.\ re-Gaussianization shape measurements. Isophotal (\dev) shapes consistently give a higher 
         amplitude by $\sim 40\%~(20\%)$ compared to re-Gaussianization shapes. The \dev\ results have been shifted 
         horizontally for clarity.
	}
         \label{fig:AI_ratio}
      \end{figure}

		Fig.~\ref{fig:lowz_wgp} shows the \wgp\ measurement for the full LOWZ sample using all 
		three shape measurement methods. 
		Isophotal shapes give the highest IA amplitude, followed by \dev\ and re-Gaussianization 
		shapes. Fig.~\ref{fig:AI_ratio} shows the amplitude trends more clearly, where we have 
		plotted the ratio of $A_I$ for isophotal and \dev\ shapes with respect to 
		re-Gaussianization shapes, for various subsamples defined by different galaxy properties. Isophotal 
		(\dev) shapes give a higher $A_I$ by $\sim40\%~(20\%)$ compared to re-Gaussianization shapes, 
		with no clear trends with galaxy properties like luminosity, color, and redshift.

	\begin{figure}
	         \centering
	         \includegraphics[width=\columnwidth]{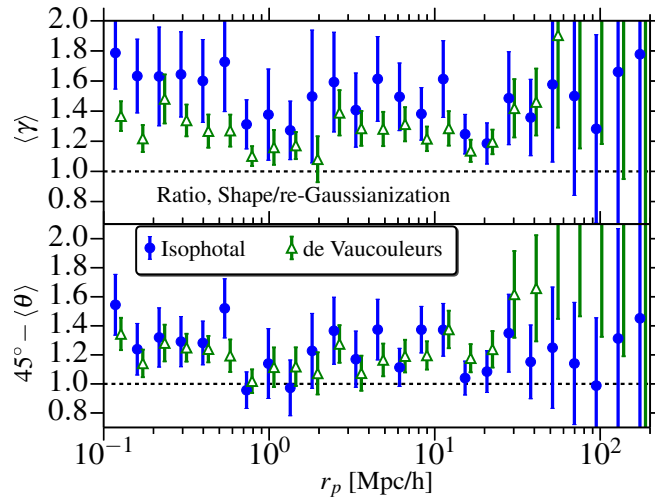}
	         \caption{Ratio of mean intrinsic shear, $\langle\gamma\rangle$ \referee{(Eq.~\ref{eqn:gamma})} and mean alignment angle, $45^\circ-\langle\theta
	         \rangle$ \referee{(Eq.~\ref{eqn:theta})} using different shape measurement methods. 
               }
	         \label{fig:lowz_theta_gamma_ratio}
	\end{figure}

		To further understand the effects of different shape measurement methods, we measure two more IA 
		estimators, $\langle\gamma\rangle$ and $\langle\theta\rangle$, defined as
		\begin{equation}\label{eqn:gamma}
				\langle \gamma \rangle =\frac{S_+D}{SD}
		\end{equation}
		\begin{equation}\label{eqn:theta}
				\langle \theta \rangle =\frac{\theta_{SD}}{SD}
		\end{equation}

		Both $\langle\gamma\rangle$ and $\langle\theta\rangle$ are calculated in a single $\Pi$ bin with $\Pi
		\in[-100,100]\mpch$.
		In the absence of intrinsic alignments, $\langle\gamma\rangle=0$ and $\langle\theta\rangle=45^\circ$, while 
		in presence of IA, $\langle\gamma\rangle>0$ and $\langle\theta\rangle
		<45^\circ$
		Fig.~\ref{fig:lowz_theta_gamma_ratio} shows the ratio of $\langle\gamma\rangle$ and $45^\circ-\langle
		\theta\rangle$ measured using isophotal and \dev\ shapes with respect to re-Gaussianization shapes. 
		$\langle\gamma\rangle$ is higher for isophotal shapes followed by \dev\ shapes, as 
		expected
		from \wgp\ results. $45^\circ-\langle\theta\rangle$ is also higher. 
	
			There are a few possible explanations for these differences in IA amplitudes with different 
		shape measurement methods. The first possibility, as discussed in Sec.~\ref{sssec:systematics_data} and 
		observed in Sec.~\ref{ssec:systematics}, is a systematic error from incorrect PSF removal. 
		Isophotal shapes are not corrected for the PSF, while \dev\ shapes are only 
		approximately corrected. As shown in Sec.~\ref{ssec:systematics}, we have a clear detection of additive 
		PSF bias in \dev\ shapes, 
		 though isophotal and re-Gaussianization shapes have much 
		lower levels of contamination. The additive bias should drop out of \wgp\ 
		calculations, under the assumption that it is not correlated with the positions of other galaxies ($\langle g
		\gamma_\text{PSF}\rangle\sim0$ in Eq.~\ref{eqn:wgp_obs}). We have not ruled out the 
		presence of multiplicative bias, which could cause an apparent change in $A_I$, but is
        extremely difficult to rule out using the data alone.  However, the
        simplest possible interpretation of how multiplicative bias should affect these results is
        not valid: if multiplicative bias is
        responsible for the lower RMS ellipticities using isophotal shapes compared to other methods
        in Fig.~\ref{fig:erms_Rcirc}, then $A_I$ should be the lowest using isophotal shapes, not the highest.
		
		\referee{Another possibility as discussed in Sec.~\ref{sssec:physical_data} is the presence of some physical 
		effect such as isophote twisting and/or ellipticity gradients. Since \wgp\ is an ellipticity-weighted measure of 
		IA, gradients in the ellipticity can also lead to higher \wgp. However, as discussed in 
		Sec.~\ref{ssec:pe}, isophotal shapes have lower \erms\ than re-gaussianization and \dev\
        shapes.  This suggests that if the galaxies have the same large-scale alignments with the
        tidal field at all radii, but an ellipticity gradient that modifies \erms, then 
		 isophotal shapes should have lowest \wgp\ amplitude, not the highest. Moreover, ellipticity gradients alone 
		cannot explain a stronger alignment angle $\langle\theta\rangle$ for isophotal shapes, as
		shown in Fig.~\ref{fig:lowz_theta_gamma_ratio}.}

\referee{The other physical effect, isophote twisting, is related to 
		the fact that outer 
		regions of galaxies may be more susceptible to the local and large-scale tidal fields and thus show 
		stronger alignments.   If this effect is important, isophotal shapes should 
		have the highest IA amplitude, followed by \dev\ and finally re-Gaussianization shapes, consistent with 
		our results. The results in Fig.~\ref{fig:lowz_theta_gamma_ratio} are also consistent
        with this physical effect.   Given that PSF-related systematics should
        cancel out of $\langle\theta\rangle$ (since the alignment is being calculated with respect
        to galaxy positions that are not aligned with respect to the PSF), and multiplicative biases
      also cannot modify the alignment angle, it is difficult to interpret that finding as anything
      other than isophote twisting.  }
	
		Finally, Fig.~\ref{fig:lowz_wpp} shows the shape-shape (\wpp) correlation functions. The \wpp\ detection using isophotal and re-
		Gaussianization shapes is not very significant, and most of the signal using \dev\ shapes can be 
		attributed to additive PSF bias as shown in Sec.~\ref{ssec:systematics}. {The detection significance 
		of our isophotal measurements is not consistent with that of \cite{Blazek2011}, who reported  a 
		high-significance detection of \wpp\ using measurements of \cite{Okumura2009} with isophotal 
		shapes.} While this could also be due to the differences in sample definition, our results in
        \pI\ suggest that even with an appropriately bright subset of LOWZ, our results may not
        agree.  Thus, we will address this discrepancy in greater detail in Sec.~\ref{ssec:comparison} after we have 
		discussed the anisotropy of IA signals, which turns out to be an important factor in this difference.
      
   \subsection{Anisotropy of IA}\label{ssec:results_anisotropy}

\begin{figure*}
	      \begin{subfigure}{\columnwidth}
	         \centering 
	         \includegraphics[width=0.95\columnwidth]{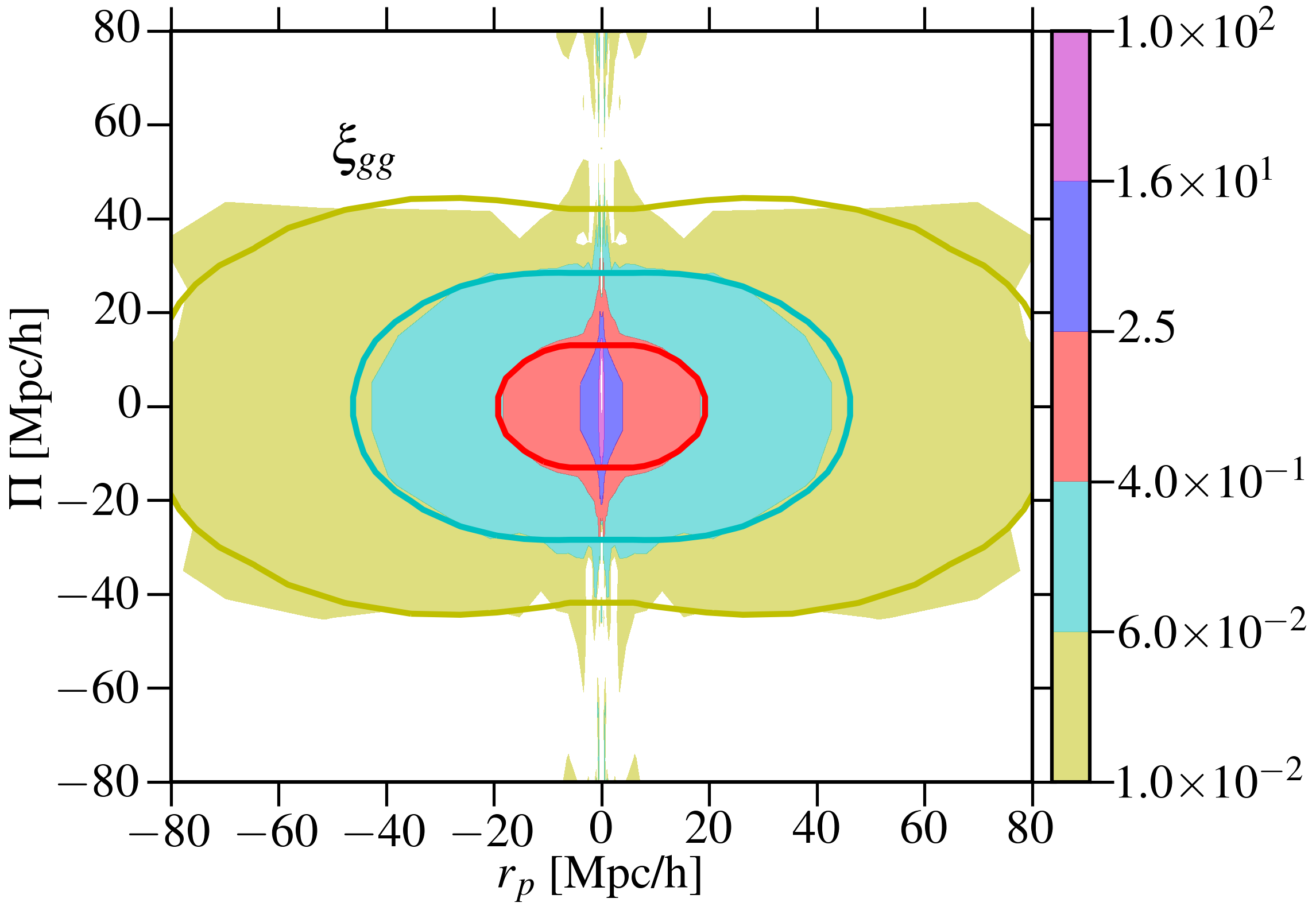} 
	         \caption{} 
	         \label{fig:wgg_contours} 
	         \end{subfigure}\hfill
	         \begin{subfigure}{\columnwidth}
	         \centering 
	         \includegraphics[width=0.9\columnwidth]{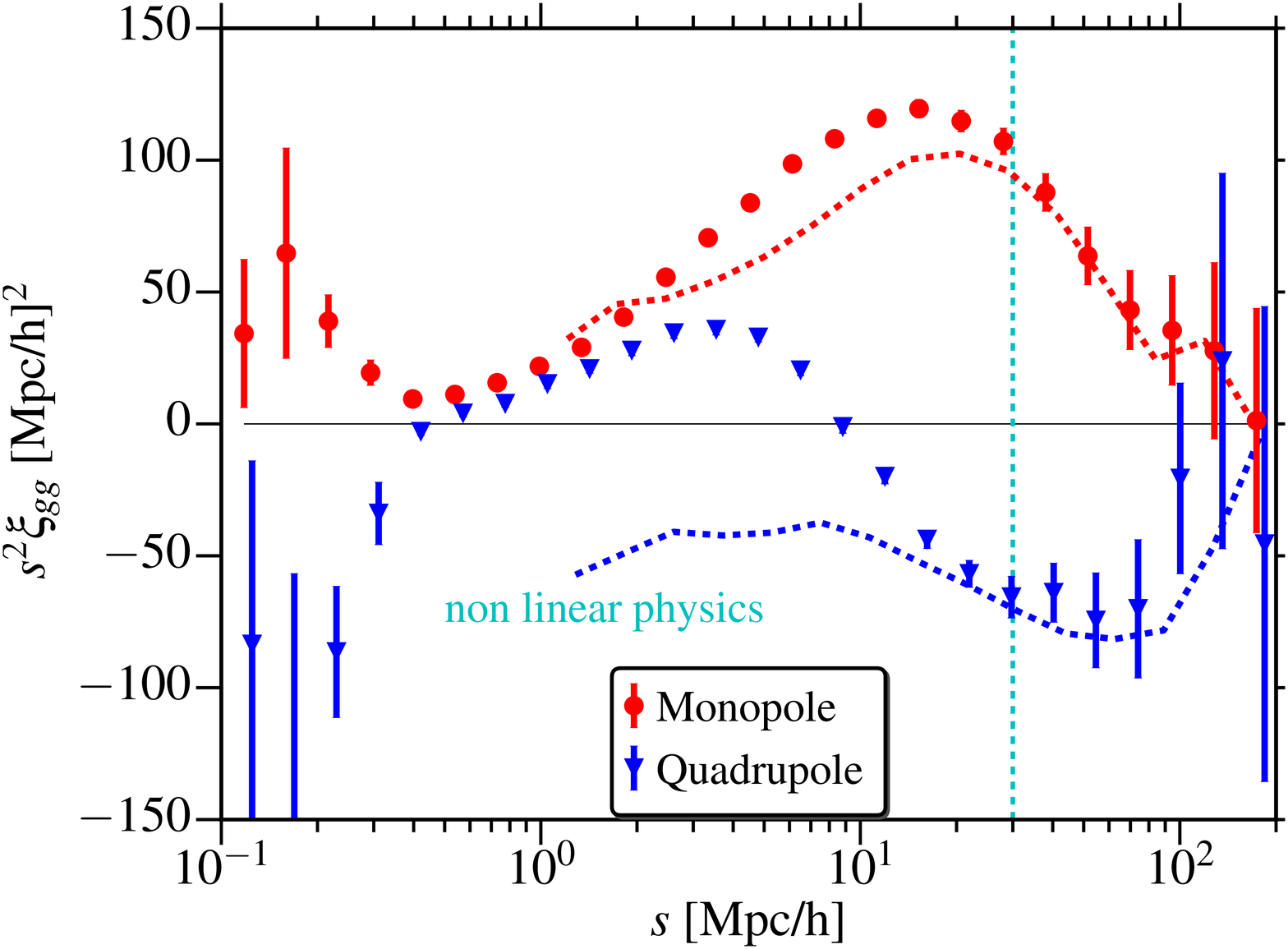} 
	         \caption{} 
	         \label{fig:wgg_multipole} 
	     \end{subfigure}

	      \begin{subfigure}{\columnwidth}
	         \centering 
	         \includegraphics[width=0.95\columnwidth]{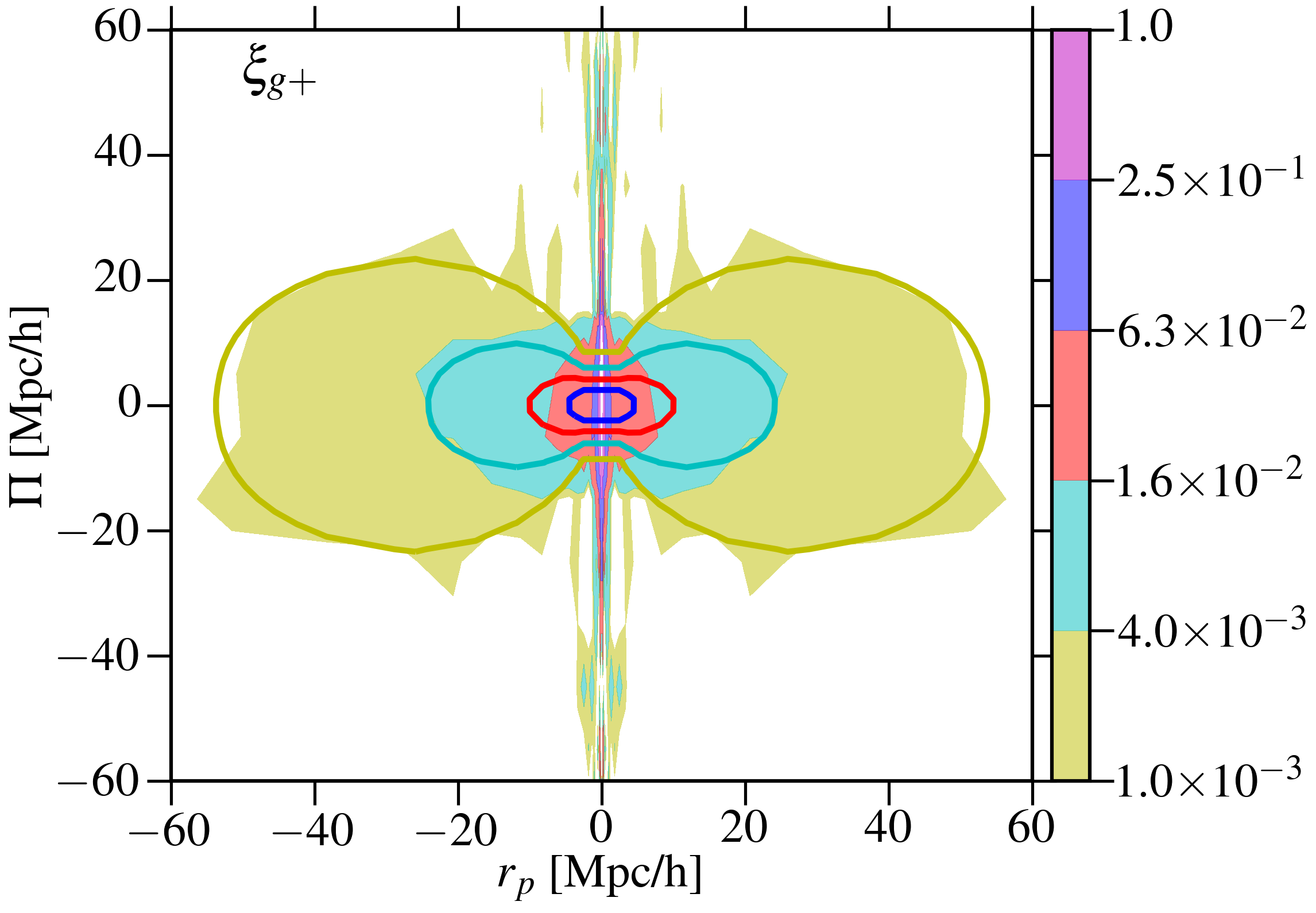} 
	         \caption{} 
	         \label{fig:wgp_contours} 
	         \end{subfigure}\hfill
	         \begin{subfigure}{\columnwidth}
	         \centering 
	         \includegraphics[width=0.9\columnwidth]{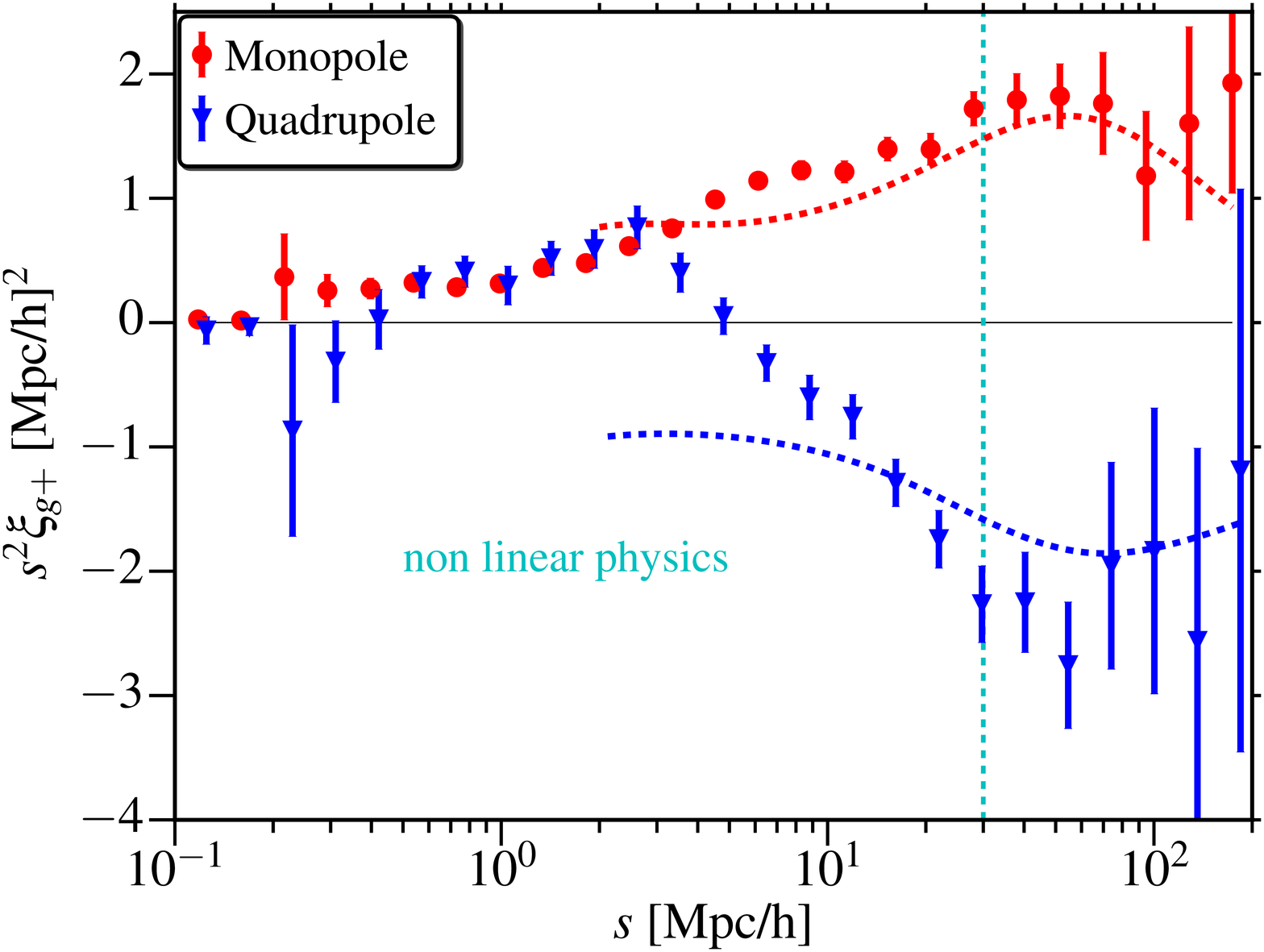} 
	         \caption{} 
	         \label{fig:wgp_multipole} 
	     \end{subfigure}
	     
	      \begin{subfigure}{\columnwidth}
	         \centering 
	         \includegraphics[width=0.95\columnwidth]{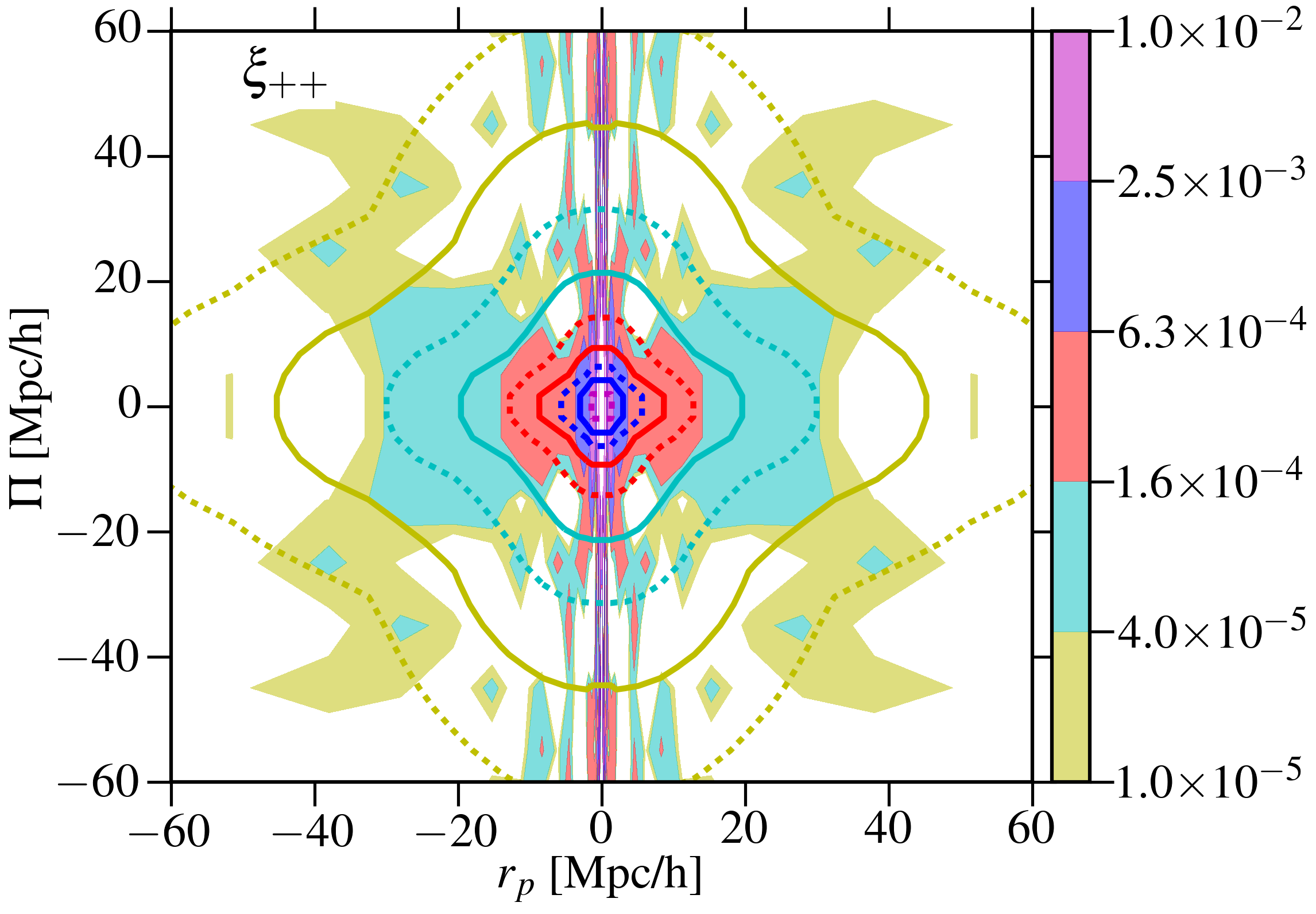} 
	         \caption{} 
	         \label{fig:wpp_contours} 
	         \end{subfigure}\hfill
	         \begin{subfigure}{\columnwidth}
	         \centering 
	         \includegraphics[width=0.9\columnwidth]{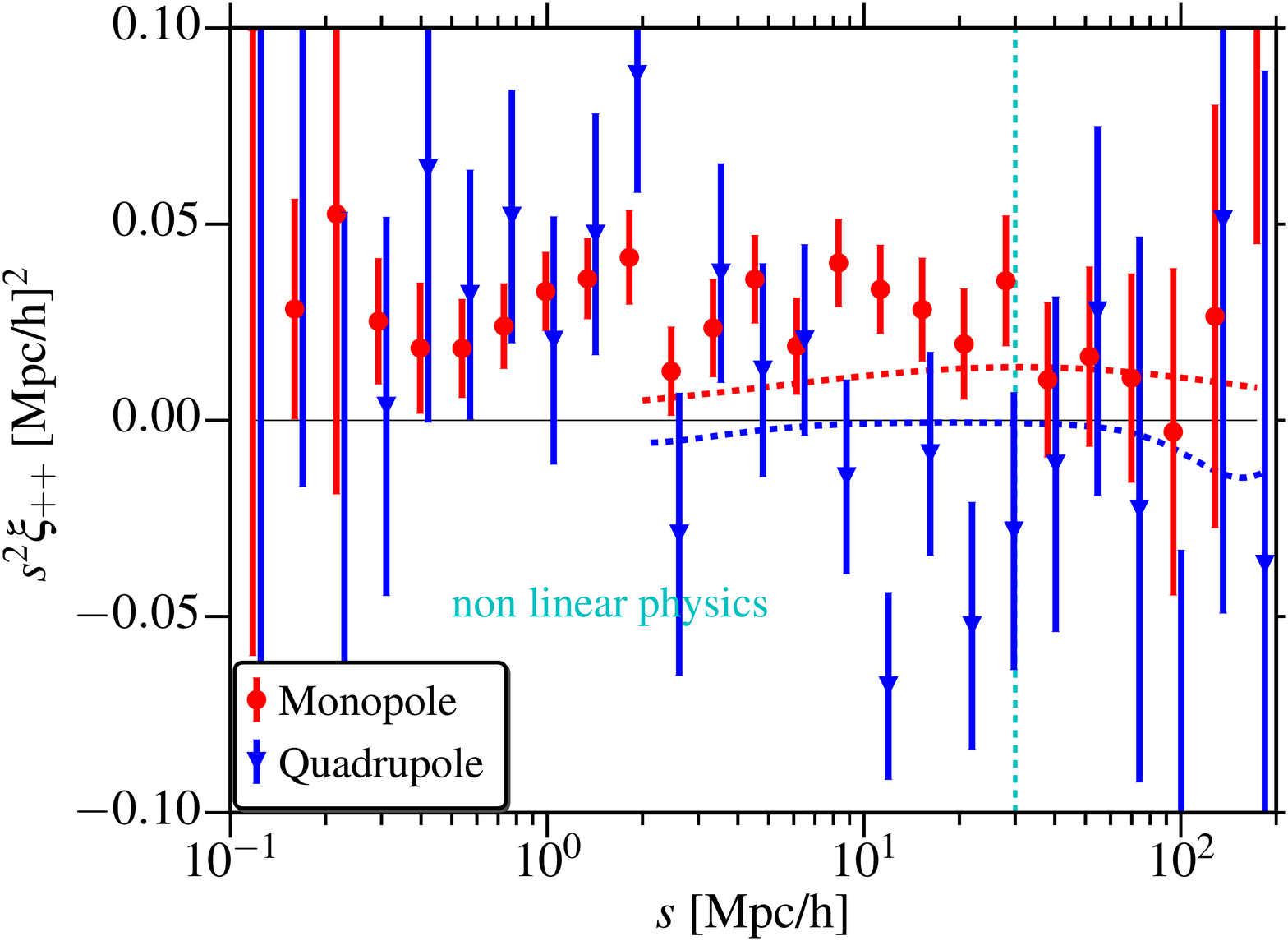} 
	         \caption{} 
	         \label{fig:wpp_multipole} 
	     \end{subfigure}

	     \caption{ \referee{The 3D galaxy-galaxy correlation function \xigg\ (top row,
             Eq.~\ref{eqn:xigg}), galaxy density-shape correlation function \xigp\ (middle 
	     row, Eq.~\ref{eqn:xigp}) and shape-shape correlation function \xipp\ (bottom row, Eq.~\ref{eqn:xipp})} as a function of $r_p,\Pi$ (left column, {reflected about $r_p=0$}) and 
	     their multipole moments (right column). All 
	     plots use isophotal shapes.  In the left column, the 
	     filled contours are showing the data, while solid lines are the theory predictions 
	     corresponding to the outer edge of the filled contours.
	     The right column shows monopole and quadrupole measurements as a 
	     function of three dimensional redshift space separation ($s=\sqrt{r_p^2+\Pi^2}$ [Mpc$/h$]). The 
	     points are  measurements from the data, while dashed lines are NLA model
         predictions. Theory predictions in both columns are from the best-fitting models to \wgg\ and \wgp, with  $f_{II}=1$ in 
         \xipp. {The dashed lines in (e) show predictions with $f_{II}=2$.}
	     The linear models with non-linear power spectrum are consistent with data for $s
	     \gtrsim30~\mpch$, below which significant deviations are expected due to non-linear RSD
         and, on even smaller scales, non-linear galaxy bias.
	     }
	     \label{fig:anisotropy} 
      \end{figure*}

   	As discussed in Sec.~\ref{ssec:anisotropy}, IA 
	 measurements suffer from anisotropy in $(r_p, \Pi)$ introduced by the projected shapes, which
     do not allow measurement of the IA signal along the line of 
	 sight. 
	 To enable a study of this effect, Fig.~\ref{fig:anisotropy} shows \xigg, \xigp, and \xipp\ as a function of $r_p$ and $\Pi$, as well as 
	 the corresponding monopole and quadrupole measurements defined in
     Sec.~\ref{ssec:anisotropy}. The top row shows the galaxy clustering 
	 measurements as well as the model predictions. Non-linear theory predictions with the Kaiser 
	 formula for RSD match the data for $r_p\gtrsim5\mpch$ and $s\gtrsim20\mpch$, 
	 though there are deviations below these scales due to nonlinear RSD and the Finger-of-God effect.
	 
\begin{figure*}
	      \begin{subfigure}{\columnwidth}
	         \centering 
	         \includegraphics[width=0.95\columnwidth]{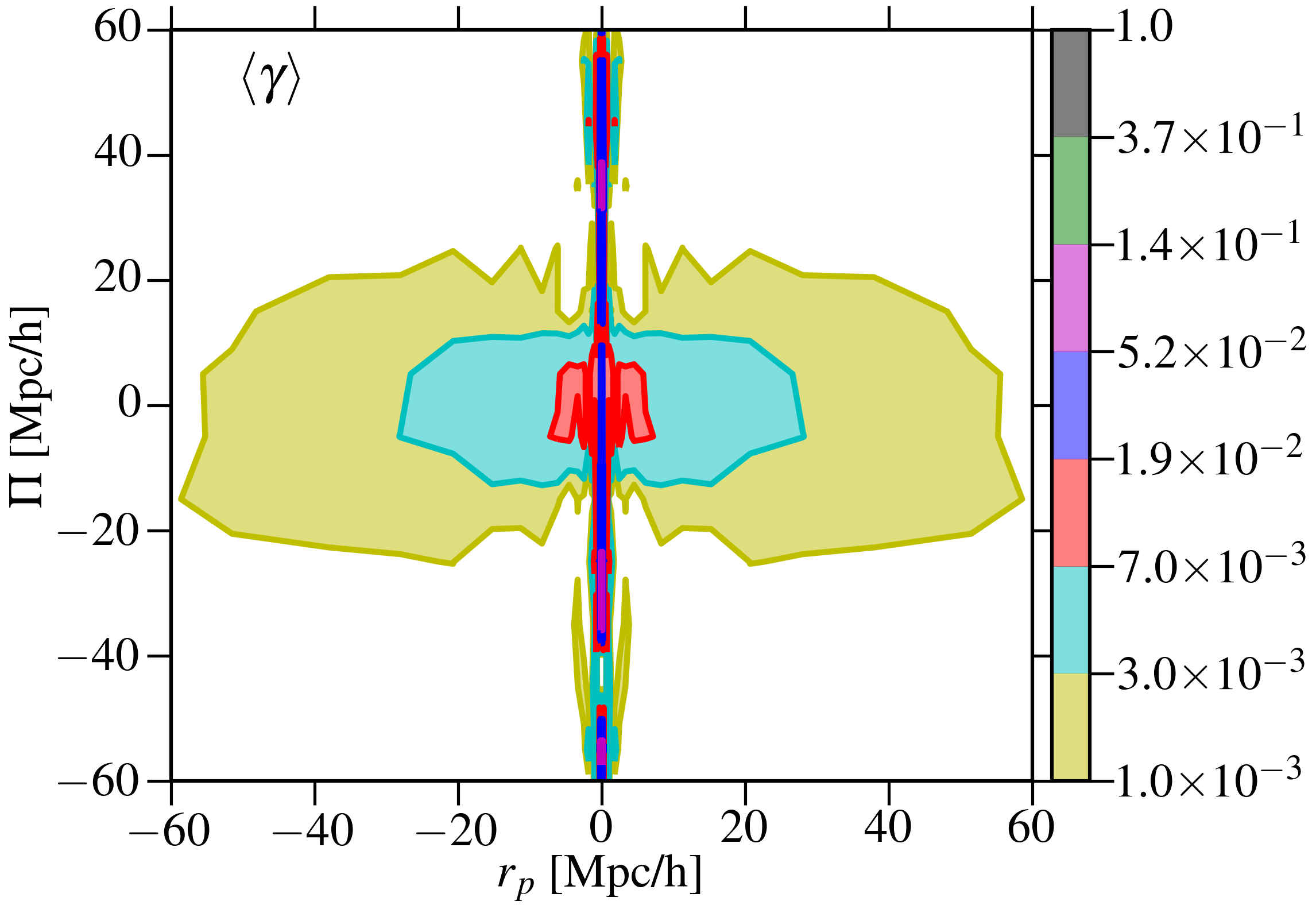} 
	         \caption{} 
	         \label{fig:wgg_contours_mb2_rsd} 
	         \end{subfigure}\hfill
	         \begin{subfigure}{\columnwidth}
	         \centering 
	         \includegraphics[width=0.9\columnwidth]{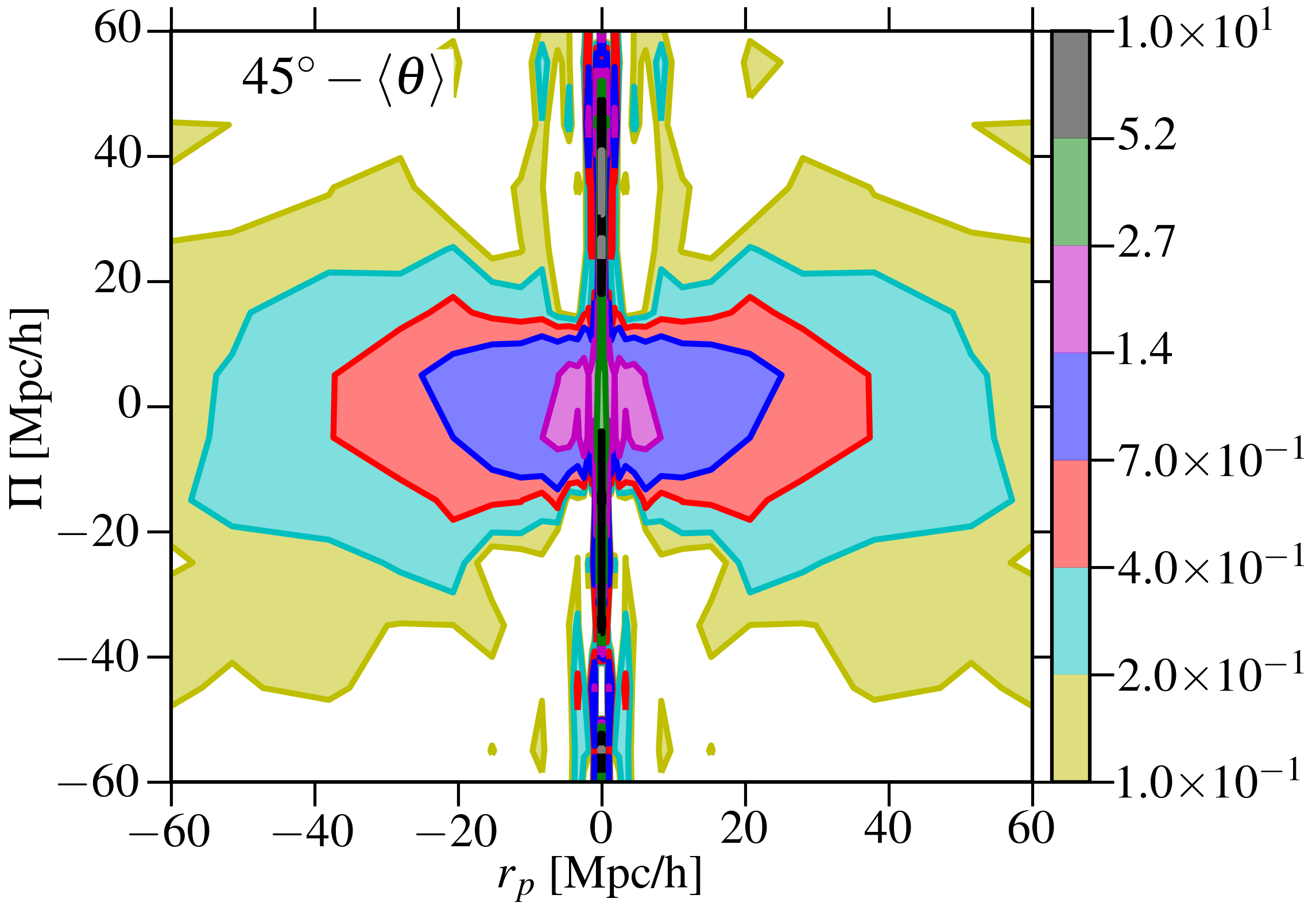} 
	         \caption{} 
	         \label{fig:theta_contours} 
	     \end{subfigure}
	     \caption{Similar to the left column of Fig.~\ref{fig:anisotropy}. (a)
           $\langle\gamma\rangle$ (Eq.~\ref{eqn:gamma}) as a function of $r_p,\Pi$. (b)
           $45^\circ-\langle\theta\rangle$ (Eq.~\ref{eqn:theta}) as a function of $r_p,\Pi$. Both
           (a) and (b) use isophotal shapes and the same binning as in Fig.~\ref{fig:anisotropy},
           but there are no theory curves on these plots.}
	     \label{fig:anisotropy_theta_gamma} 
      \end{figure*}

	 The middle row in Fig.~\ref{fig:anisotropy} shows \xigp\ measurements using isophotal shapes. 
	 Again, the NLA model fits the data well for $r_p\gtrsim5\mpch$ and $s\gtrsim20\mpch$ with 
	 deviations at small scales due to nonlinear RSD {and clustering}. The strong anisotropy introduced by the 
	 projected shapes is clearly visible in the shape of the peanut-shaped contours, where the signal 
	 drops off quickly with $\Pi$. 
	 The NLA model incorporates this anisotropy and is consistent with the data. 
{The redshift-space structure of $\langle\gamma\rangle$ 
	 and $\langle\theta\rangle$ (shown in Fig.~\ref{fig:anisotropy_theta_gamma})
	 is very similar to that of \xigp.
}	
 
\begin{figure}
	         \centering
	         \includegraphics[width=\columnwidth]{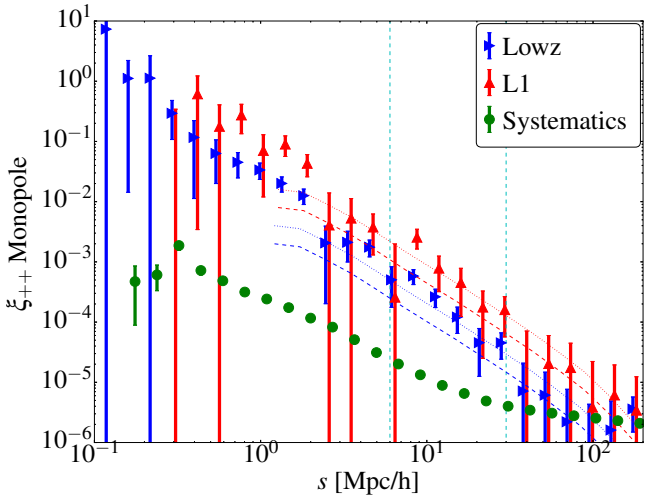}
	         \caption{
	         \referee{The 3D shape-shape correlation function \xipp\ (Eq.~\ref{eqn:xipp})} monopole moment using isophotal shapes for LOWZ (blue) and the $L_1$ (red) subsample, along with 
	         theory predictions from the best-fitting NLA model to \wgp\ with $f_{II}=1$ (dashed lines) and
             $f_{II}=2$ (dotted lines). Green points also show the monopole of the expected additive PSF
             contamination based on the analysis in Sec.~\ref{ssec:systematics}, 
             $\xipp^\text{sys}\approx A_\text{PSF,isoph}^2\times \xipp^\text{PSF-PSF}$. Vertical lines at $s=6$ and $30\mpch$ 
             show the limits used to calculate $\chi^2$ values to assess the goodness of the  model.
               }
	         \label{fig:isophotal_monopole}
	\end{figure}

 {
	 The bottom row in Fig.~\ref{fig:anisotropy} shows \xipp\ measurements using isophotal shapes.
	 To display NLA model predictions, we use the best-fitting parameters from fitting \wgp, with $f_{II}=1$ (solid lines) 
	 and $f_{II}=2$ (dashed lines). The two-dimensional contours in Fig.~\ref{fig:wpp_contours}
     suggest that the data prefer the 
	 model with $f_{II}=2$. However, these contours are quite 
	 noisy, so Fig.~\ref{fig:wpp_contours} is not a  reliable test of the validity of the model. 
	 In Fig.~\ref{fig:wpp_multipole}, we show a clear detection of the monopole for \xipp, with the data
     again preferring a 
	 higher amplitude than predicted by NLA model ($f_{II}>1$). This discrepancy could either be
     from the effects of non-linear physics that is not included in the NLA model
     \citep{Blazek2015}, or from additive  PSF contamination.
	  Even though additive PSF contamination was shown to be low for 
	 isophotal shapes ($|A_{PSF}|\sim0.05$), the contamination in \xipp\ could still be strong
     enough to increase the observed \xipp\ 
	 amplitude. In Fig.~\ref{fig:isophotal_monopole}, we show the \xipp\ monopole for the full LOWZ sample and the 
	 brightest subsample, $L_1$, along with the predicted PSF contamination. For large scales, $s>30$\mpch, the PSF 
	 contamination alone can account for the observed \xipp\ signal for the full LOWZ sample. At smaller scales, the PSF 
	 effects are 
	 subdominant, which make it unlikely to be responsible for the higher than predicted amplitude. Both LOWZ and $L_1$ 
	 samples prefer a higher amplitude; based on $\Delta \chi^2$, the $p-$value$=0.01$ ($0.15$) for
     LOWZ ($L_1$).  However, the 
	 $f_{II}=1$ model is not 
	 ruled out by either sample, with a $p$-value of $0.07$ ($0.44$) for LOWZ ($L_1$). With 
	 PSF 
	 effects being subdominant, we conclude that the discrepancy between NLA and data is primarily
     due to non-linear physics (e.g.\ non-linear clustering, non-linear RSD), which become important for scales $s<30\mpch$ and 
	 generally increase the amplitude \citep{Blazek2015}. 
	 }
	 	 
	 The \xipp\ quadrupole moment in both the data and the NLA model prediction is close to zero. 
     \xipp\ is relatively more isotropic {\citep[][]{Croft2000}} than \xigp\ as some large-scale 
	 modes along the line-of-sight can introduce similar IA in projected galaxy shapes with larger separation along 
	 $\Pi$. These correlations show up in the shape-shape correlation function but not in the
     density-shape correlations. Mathematically, these terms are sourced by the $J_0$ term combined with $(1-\mu^2)^2$ 
     (see Eq.~\ref{eqn:xipp}) which is more extended along $\Pi$, while the $J_2$ and $J_4$ terms are more extended 
     along $r_p$. The combination of $J_0$ and $J_4$ terms makes \xipp\ relatively more isotropic
     than \xigp.

\begin{figure*}
	      \begin{subfigure}{\columnwidth}
	         \centering 
	         \includegraphics[width=0.95\columnwidth]{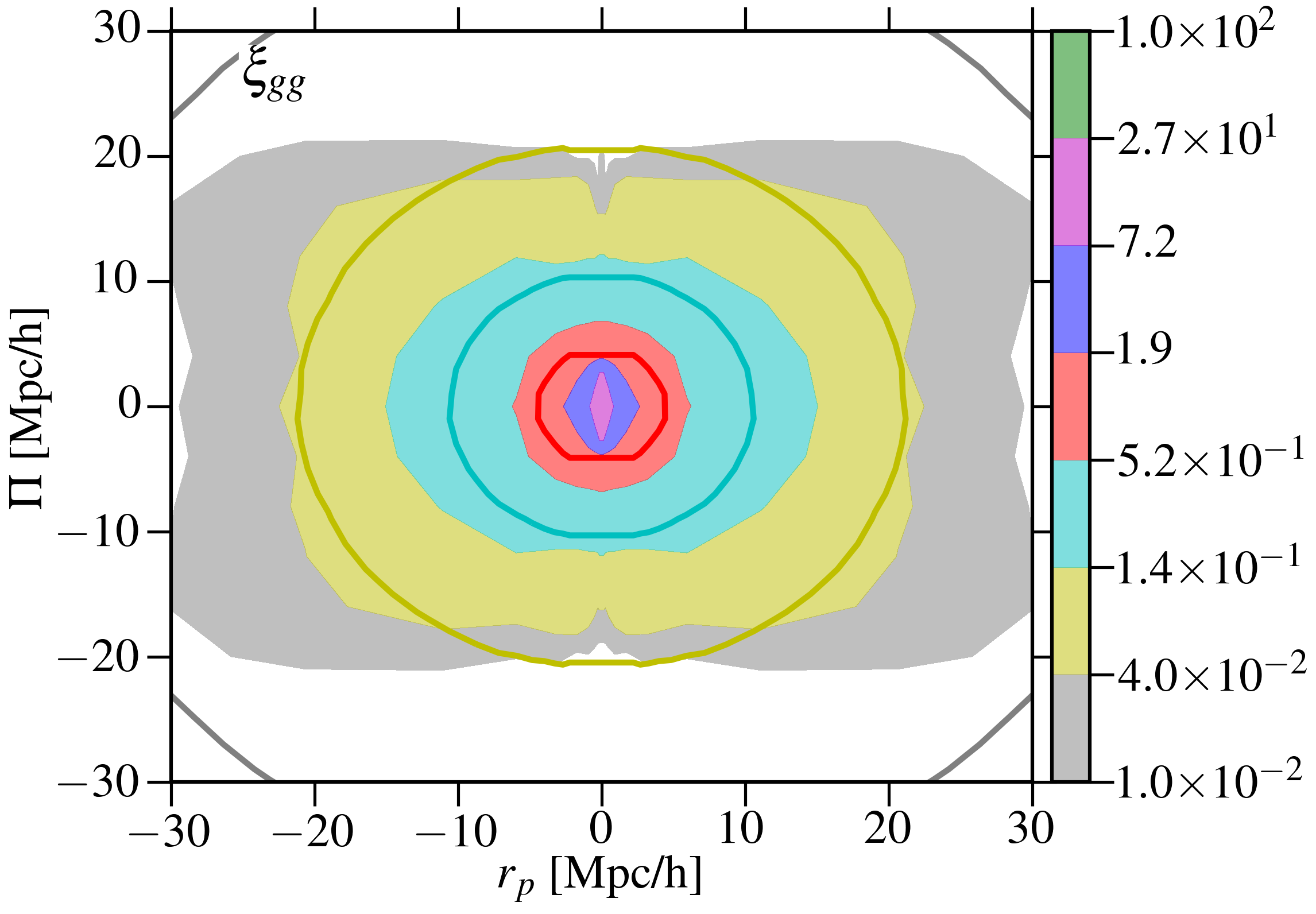} 
	         \caption{} 
	         \label{fig:wgg_contours_mb2} 
	         \end{subfigure}\hfill
	         \begin{subfigure}{\columnwidth}
	         \centering 
	         \includegraphics[width=0.9\columnwidth]{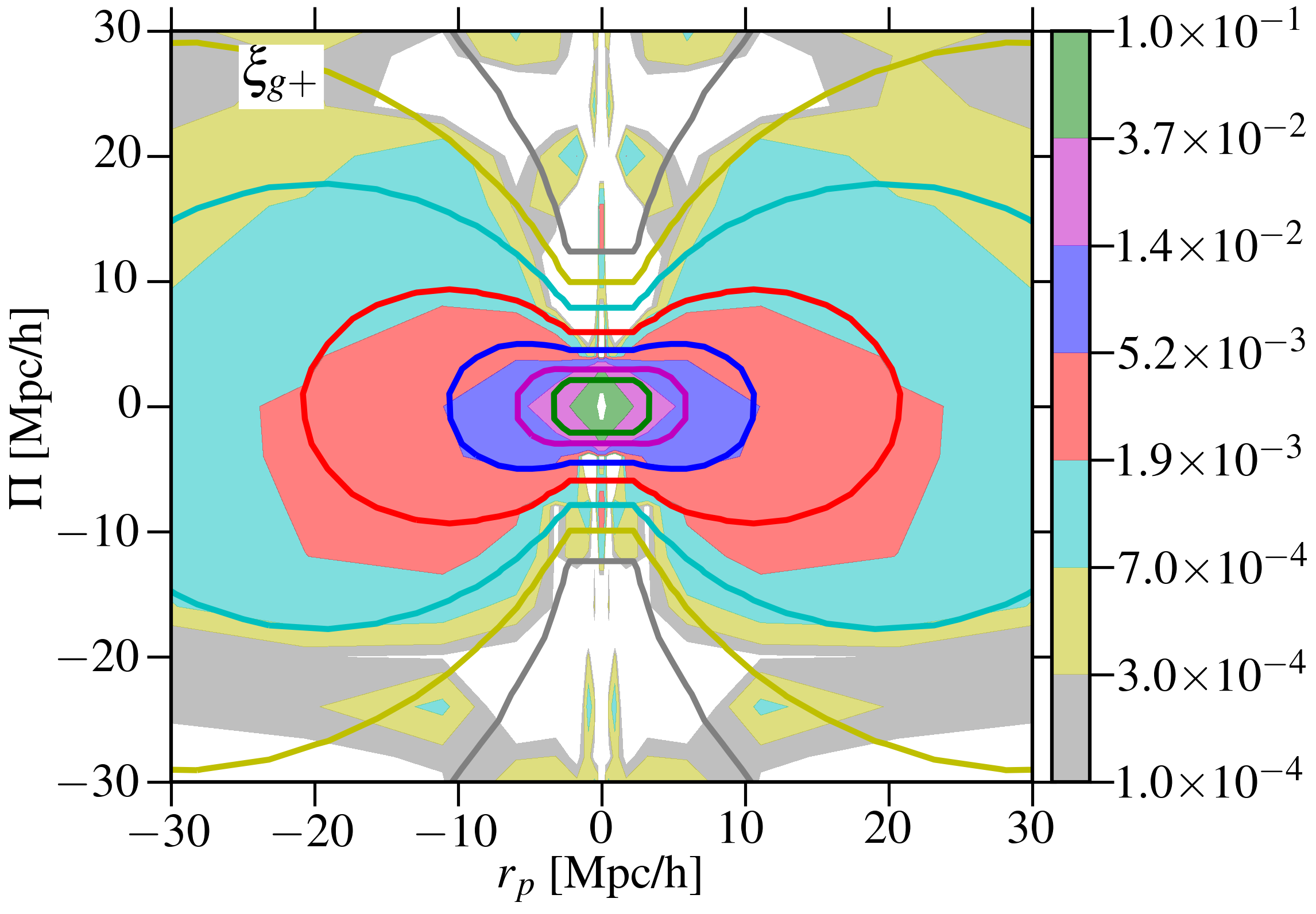} 
	         \caption{} 
	         \label{fig:wgp_contours_mb2} 
	     \end{subfigure}
	      \begin{subfigure}{\columnwidth}
	         \centering 
	         \includegraphics[width=0.95\columnwidth]{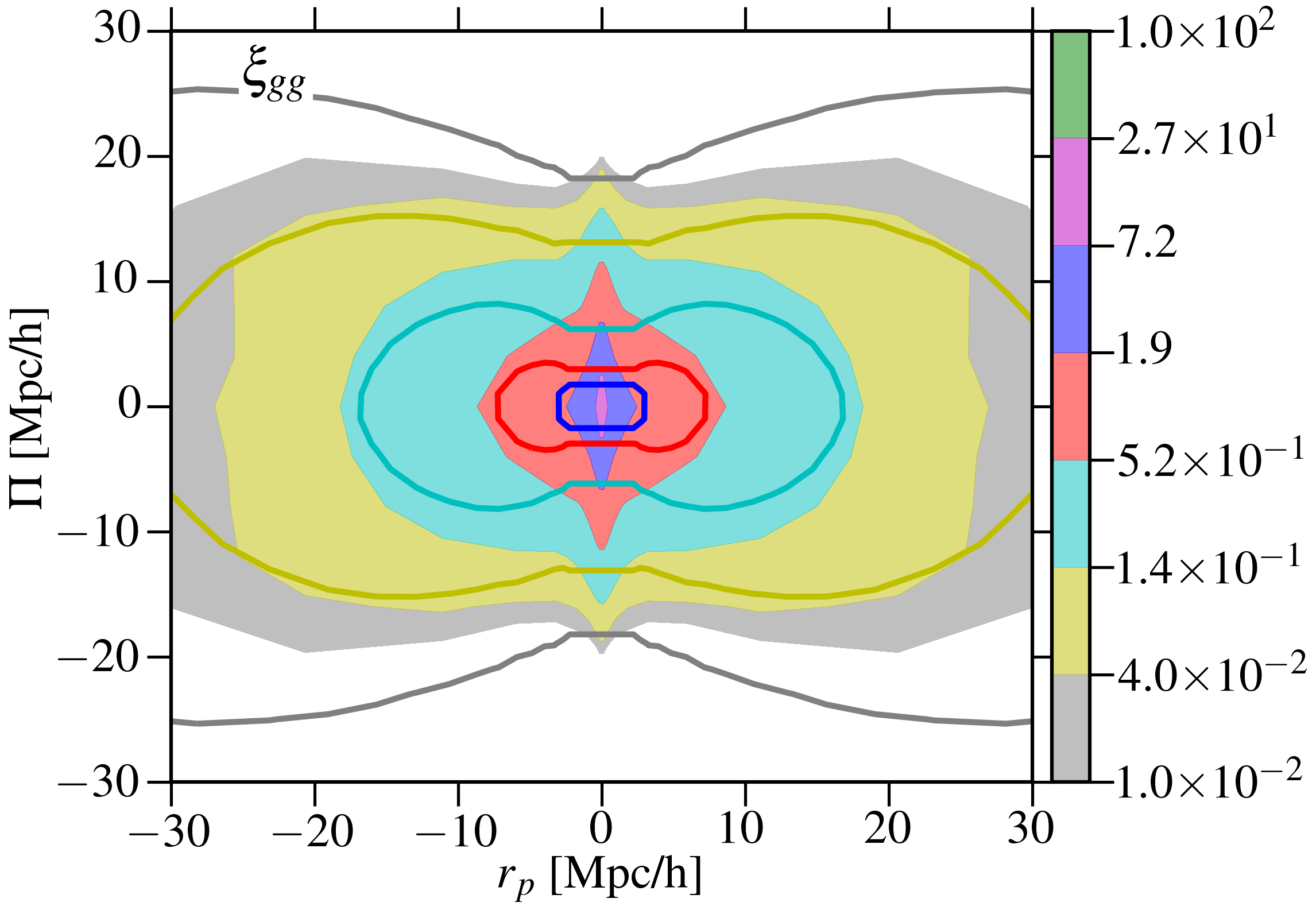} 
	         \caption{} 
	         \label{fig:wgg_contours_mb2_rsd} 
	         \end{subfigure}\hfill
	         \begin{subfigure}{\columnwidth}
	         \centering 
	         \includegraphics[width=0.9\columnwidth]{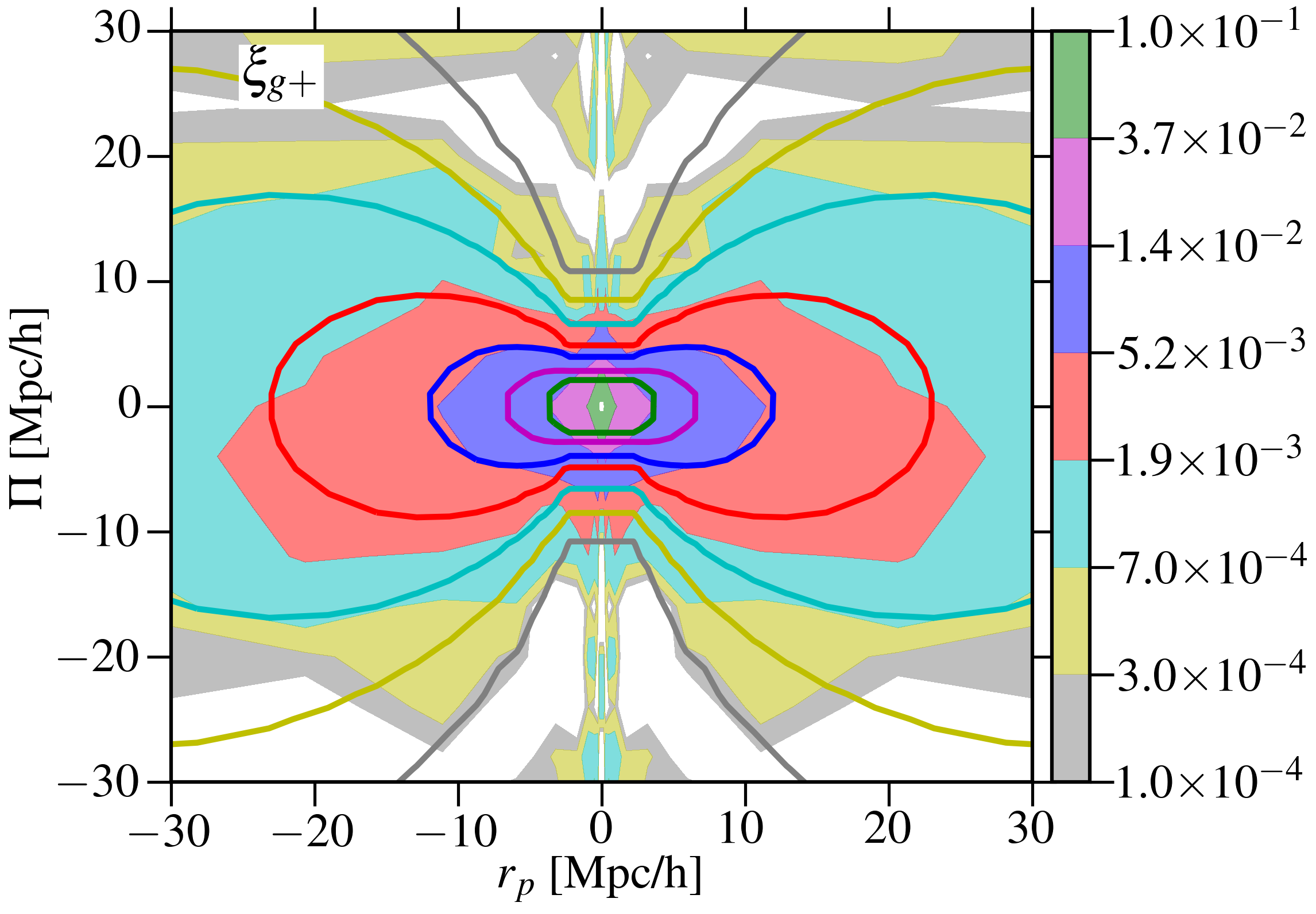} 
	         \caption{} 
	         \label{fig:wgp_contours_mb2_rsd} 
	     \end{subfigure}
	     \caption{\referee{The 3D galaxy-galaxy correlation function \xigg\ (Eq.~\ref{eqn:xigg}) and galaxy
           density-shape correlation function \xigp\ (Eq.~\ref{eqn:xigp})} contour plots from the MB-II simulation, similar to Fig.~\ref{fig:wgg_contours} and 
	     Fig.~\ref{fig:wgp_contours}. The top row shows the signal and model without RSD, while the
         bottom row includes  RSD. We used linear bias $b=0.8$ and $A_I=6$ for the theory
         predictions. {Note that the small scale behavior in (a,b) is an artifact from
           binning, with logarithmic bins in $r_p$ and linear bins in $\Pi$ with $d\Pi=5\mpch$. }
	     }
	     \label{fig:anisotropy_mb2} 
      \end{figure*}

As a final exploration of the anisotropy of intrinsic alignments, Fig.~\ref{fig:anisotropy_mb2}
shows \xigg\ and \xigp\ measured using the MassiveBlack-II (MB-II) cosmological hydrodynamic simulation \citep{Khandai2015} at 
	$z=0.3$, along with NLA model predictions. The top row shows the signal without RSD, along with model predictions 
	with $\beta=0$. Bottom row shows both signal and model with RSD effects included. The effects of
    RSD are clearly visible in \xigg. On the other hand, the variations in \xigp\ from top to bottom
    row  are far less pronounced, consistent with our findings that \xigp\ is not affect by RSD to
    first order, $\beta_{\gamma}=0$ (see \pI\ for derivation).  Qualitatively, the simulation 
	results produce the features seen in the data and show good agreement with the model within the 
	limitations of validity of the 
	model (which does not include nonlinear RSD or nonlinear galaxy bias, resulting in small-scale discrepancies). At large scales ($r\gtrsim20\mpch$) cosmic variance plays an important role due to the
	limited size of the simulation box ($100\mpch$), making it difficult to compare the model and
    simulation data at
    these scales.
   

	\subsection{Comparison with other studies}\label{ssec:comparison}

		As discussed in \pI\ and Sec.~\ref{ssec:results_shapes}, there is an apparent discrepancy in detection 
		significance of our \wpp\ measurements with those of \cite{Blazek2011}. \cite{Blazek2011} 
		used a projection of 3D results from \cite{Okumura2009}, who measured \xipp\ ($C_{11}$ 
		in their notation) as function of redshift-space separation $s$ using isophotal shapes. We hereafter use $C_{11}$
        to refer to $\xipp(s)$, to distinguish it from our measurement of $\xipp(r_p,
        \Pi)$. Assuming isotropy of $\xipp(s)$, 
		\cite{Blazek2011} projected the $C_{11}$ measurement onto $(r_p, \Pi)$ space in order to 
		calculate \wpp. Their error estimates were based on 
		generating 1000 random realizations of $C_{11}$, assuming Gaussian and independent 
		errors. To test whether any parts of this procedure (either the signal or error estimation)
        could lead to differences in the estimated $S/N$ compared to our procedure, we 
		repeat their analysis for our LOWZ sample, using isophotal shapes.

\begin{figure}
	         \centering
	         \includegraphics[width=\columnwidth]{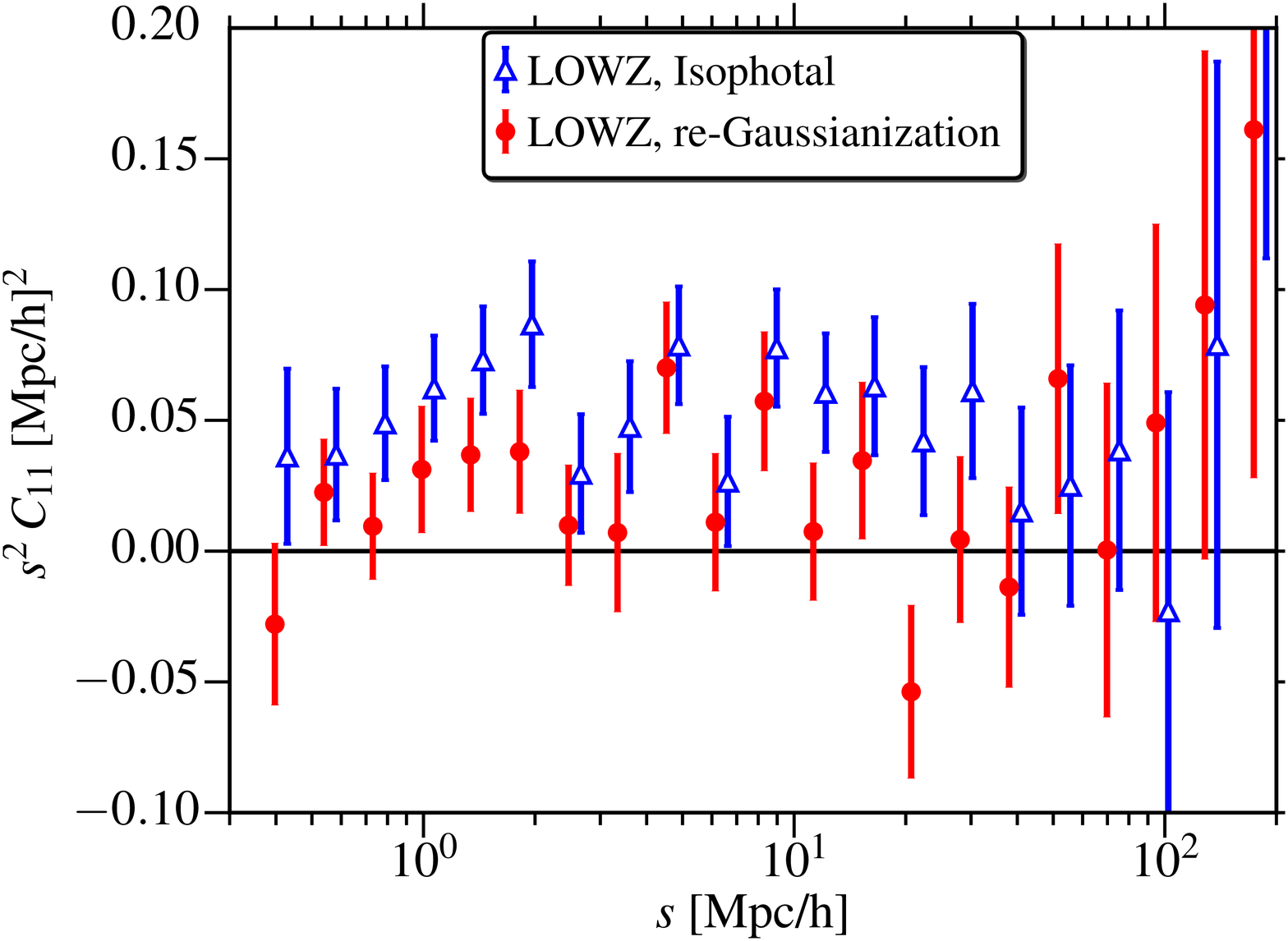}
	         \caption{ The 3D shape-shape correlation function as a function of redshift-space
               separation $s$, $C_{11}(s)$ {(or $\xipp(s)$)}, for the entire LOWZ sample, using 
               re-Gaussianization and isophotal shapes. 
	         }
		 \label{fig:lowz_iso_c11}
     	      \end{figure}

		Fig.~\ref{fig:lowz_iso_c11} shows our measurement of $C_{11}(s)$. 
		With isophotal shapes, we have a significant detection of shape-shape correlations (as do 
	\citealt{Okumura2009} for LRGs), whereas the detection is not significant using
    re-Gaussianization shapes. We do not attempt a quantitative 
		comparison with their results, since differences in sample selection complicate the comparison.

\begin{figure}
	         \centering
	         \includegraphics[width=\columnwidth]{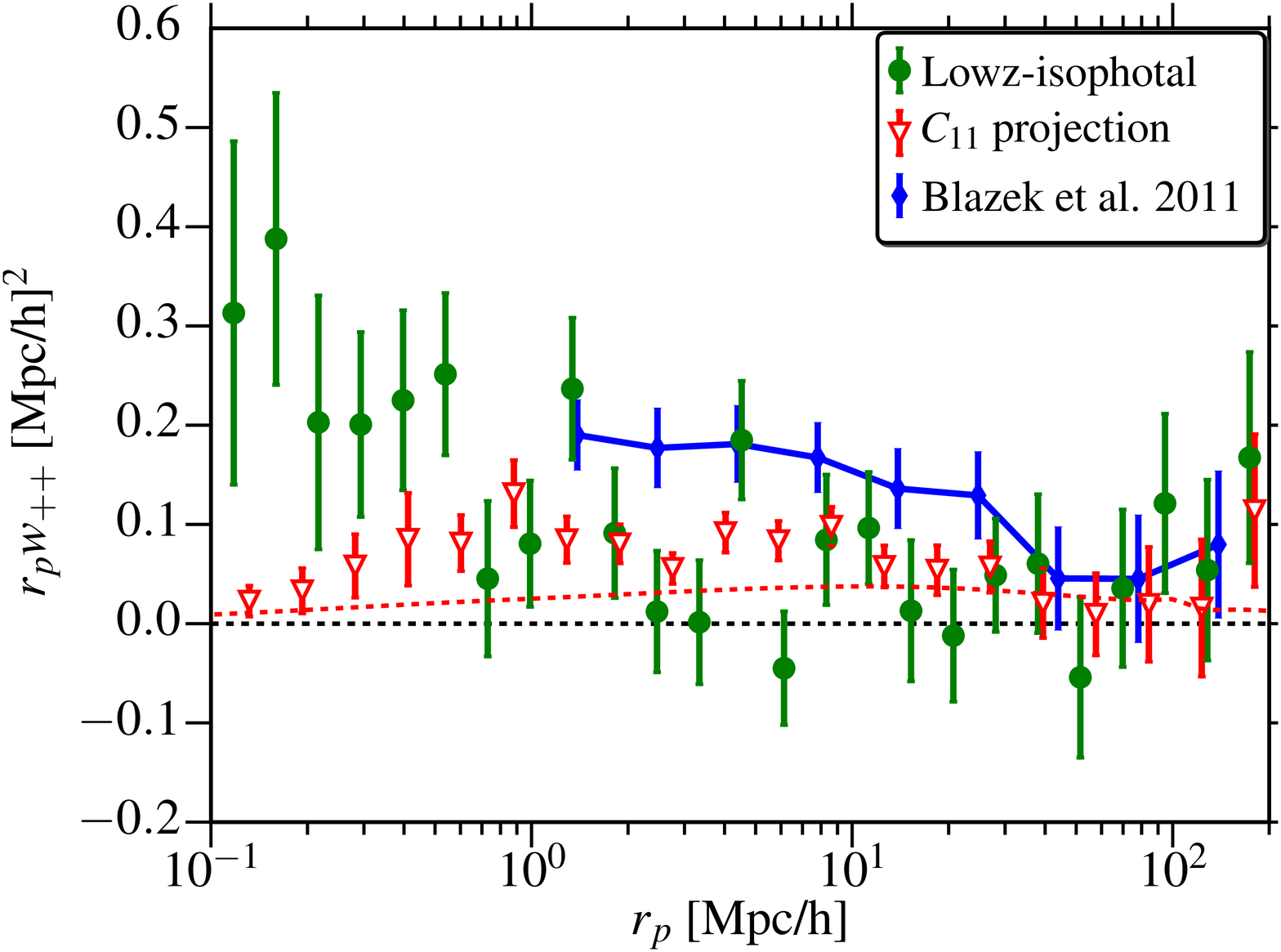}
	         \caption{Comparison of \wpp\ measured in different ways. The green points show the signal 
	         after measuring \xipp\ in $(r_p, \Pi)$ space and projecting along $\Pi$ using our
             standard methodology, while the red points show the signal obtained by 
	         projecting $C_{11}(s)$ following the approach of \protect\cite{Blazek2011}. The
             flatness of \wpp\ using the second approach at $r_p\lesssim 1\mpch$ is due to the choice of bin sizes 
	         ($d\Pi=1\mpch$); see text for details. The signal for brighter LRGs from \protect\cite{Blazek2011} 
	         signal is plotted in green for reference, but should not be compared in detail with our
             results due to differences in sample definition.
	         }
		 \label{fig:wpp_projection}
     	      \end{figure}

		 Using this $C_{11}(s)$ measurement with isophotal shapes, we repeat the procedure of \cite{Blazek2011} to 
		 get $\wpp(r_p)$. We project the 
		 $C_{11}$ measurement onto the $(r_p, \Pi)$ plane, using 200 $\Pi$ bins with $\Pi \in [-100,100]\mpch$ and 
		 $d\Pi=1\mpch$.
		 Fig.~\ref{fig:wpp_projection} compares the \wpp\ obtained using different 
		 methods. The \wpp\ for the LOWZ sample calculated using the \cite{Blazek2011} method of
	         projecting $C_{11}(s)$ does have a higher detection 
		 significance compared to the \wpp\ using our $\xipp(r_p, \Pi)$ projection. This result is most likely 
		 due to the fact that $C_{11}$ is 
{effectively the monopole term,
		 which we detect with high significance (Fig.~\ref{fig:wpp_multipole}). $C_{11}$ and the monopole have different 
		 amplitudes due to an additional $(2l+1)/2$ factor in the multipole moments (see Eq.~\ref{eqn:multipole}). The 
		 differences in $S/N$ between $C_{11}$ and \wpp\ primarily come from the different
         sensitivity of the two estimators to  the 
		 large $\Pi$ regions that contribute little signal but significant noise. In accordance with
         this explanation, we find that using $\Pi_\text{max}\sim30\mpch$ when 
		 computing \wpp\ does give a statistically significant measurement (not shown).
		 } 
		 To compare error estimates, we repeat the \cite{Blazek2011} analysis but obtain 
		 errors using 100 jackknife regions instead of using 1000 random realizations of $C_{11}(s)$. The errors 
		 obtained using
		 both methods are consistent. Note that since the LRG sample used by \cite{Okumura2009} is brighter 
		 than the LOWZ sample, the amplitude for the \cite{Blazek2011} \wpp\ measurement is expected
	         to be higher than for LOWZ. 
		Also, the choice of $d\Pi=1~\mpch$ leads to an approximately constant \wpp\ for $r_p<1\mpch$ 
		 since all bins are dominated by $C_{11}$ values from $s\sim 1\mpch$, given that the signal drops 
		 exponentially with increasing $s$ and bins with large $\Pi$ do not contribute much. Hence, the apparently 
		 constant \wpp\ at $r_p<1\mpch$ in Fig.~\ref{fig:wpp_projection} is not physical.

		{In Fig.~\ref{fig:wpp_projection} we also show the theory prediction from the
          best-fitting NLA model to \wgp. As in Fig.~\ref{fig:isophotal_monopole}, the data prefer a
          higher amplitude than predicted by the NLA model,  likely due to the effects of non-linear
          physics beyond the NLA model. This result is inconsistent with the findings of \cite{Blazek2011}, who found consistent IA amplitudes from both \wgp\ and \wpp. Their sample, however, is similar to our $L_1$ sample, which as shown in Fig.~\ref{fig:isophotal_monopole} is consistent with $f_{II}=1$. 
		}

  To summarize, we have investigated three possible reasons for the apparent difference in \wpp\
detection significance in \cite{Blazek2011} compared to \pI. { We find that the primary causes are
(a) use of isophotal rather than re-Gaussianization shapes (increased signal), and (b) the projection of $C_{11}(s)$
rather than $\xipp(r_p, \Pi)$ (which ignores noisy higher multipoles), but conclude that their error estimate procedure is in agreement 
with the jackknife procedure.  Hence the different $S/N$ in \wpp\ arises from differences in 
signal estimation (the first of which increases the signal, the second of which  lowers the noise). }


Several other studies have used \dev\ and isophotal shapes to measure IA using the bright, nearby
SDSS Main sample and the fainter, more distant BOSS 
CMASS samples \citep{Hao2011,Li2013,Zhang2013}. These samples have different redshifts and galaxy properties 
compared to LOWZ. As a result, the galaxy detection have different $S/N$ and resolution compared to
the PSF, which should modify the observational 
systematics in galaxy shapes. Thus, our results about differences in systematics and in the IA
amplitude using these different shape measurement methods cannot be used to make definitive
statements about systematic effects in those studies. 

\section{Conclusions}

In this work, we have studied SDSS-III BOSS LOWZ galaxy shapes and intrinsic alignments using three different shape 
measurement methods: re-Gaussianization (PSF-corrected, weighted towards inner regions of galaxies),
isophotal (based on the shape of a low surface brightness isophote at large radius, not PSF-corrected), and \dev\
shapes (from fitting a \dev\ model to the light profile using an approximate PSF model, using a
grid-based procedure that results in some quantization of model parameters). 
Different shape measurement methods give a different ellipticity for galaxies, which in the absence
of systematic error implies a radial gradient in the galaxy shapes, with the shapes becoming rounder
on average 
at large radius.  These variations in ellipticity seem to depend on the galaxy environment, with
brightest group galaxies (BGGs) actually becoming 
more elliptical with radius but satellites and field galaxies becoming rounder.
The overall sign of the ellipticity gradients is consistent with hydrodynamic simulations \citep{Tenneti2015b}
and the environment trends are consistent with those seen using a small sample of elliptical
galaxies \citep{Tullio1978,Tullio1979}. We caution, however,
that the isophotal shapes on which these conclusions rest 
are not corrected for the PSF.  Biases from the PSF {(of which we see hints in Fig.~\ref{fig:erms_Rcirc} and~\ref{fig:isophotal_monopole})} and other observational systematics (e.g.,
contamination from light in nearby galaxies around BGGs) can alter these conclusions.

 {Tests for systematics do not reveal  a level of multiplicative or
  additive bias in either re-Gaussianization or isophotal shapes that could significantly 
  change the measured intrinsic alignment statistics for LOWZ galaxies on scales up to a few tens of
  Mpc.}
In contrast, tests for additive systematic errors in the shape-shape correlation functions reveal that the \dev\ shapes are
significantly affected by additive PSF bias.  The magnitude of the systematic in the \dev\ shapes is consistent
with $\sim 30$\% of the PSF anisotropy leaking into the galaxy shapes as an additive term.

A comparison of the density-shape correlations (\wgp) using the different shape measurement methods
revealed that isophotal (\dev) shapes give $\sim 40~(20)\%$ higher NLA model amplitude, $A_I$, 
compared to re-Gaussianization shapes. Since isophotal shapes are slightly rounder on average, this finding
cannot be easily explained in terms of multiplicative bias.  These differences
in the IA results may imply isophote twisting of galaxy shapes to make the outer regions more aligned with the tidal field, 
consistent with theoretical predictions \citep{Kormendy1982,Kuhlen2007}.  We emphasize that our
conclusions may not carry over to studies that use significantly fainter or less well-resolved
galaxy samples (e.g., the BOSS CMASS sample), where systematic errors are likely to be more
important.  Use of a suite of systematics tests as shown in this paper can be helpful to reveal
problems; however, the issue of multiplicative bias will be difficult to completely resolve from the data alone.

We also studied the anisotropy of IA signal as a function of $r_p$ and $\Pi$, finding that NLA model
predictions and those from hydrodynamic simulations are consistent with the observations.  The
projection factor from 3D to 2D shapes is the dominant source of anisotropy in the 3D density-shape
correlations \xigp, resulting in peanut-shaped contours in the $(r_p, \Pi)$ plane.

Finally, we investigated the difference in shape-shape intrinsic alignments (\wpp) detections in
\cite{Blazek2011} vs.\ \pI, and identified two significant sources of differences: use of isophotal
rather than re-Gaussianization shapes, and estimation of the signal via projection of $C_{11}(s)$
under the assumption of isotropy in the $(r_p, \Pi)$ plane rather than via direct projection of
$\xipp(r_p, \Pi)$. 

Our results have implications for intrinsic alignments forecasting, i.e., the prediction of IA
contamination in weak lensing measurements, and for intrinsic alignments mitigation with future
surveys.  First, in the case of forecasting, the relevant point is the large variation (up to 40\%)
in the NLA model amplitude that is inferred from \wgp\ measurements using different shape
measurement methods.  What is the appropriate amplitude to use for forecasts?   We argue that the
relevant one to use depends on the shape measurement method used for estimation of shear in the weak
lensing survey for which forecasts are being done.  If using a more centrally-weighted method, our
results suggest that a lower IA amplitude is more appropriate and will give more accurate
forecasts.

Second, mitigation of IA may involve joint modeling of IA and 
lensing in measurements of shape-shape, galaxy-shape, and galaxy-galaxy correlation functions
\citep[e.g.,][]{2010A&A...523A...1J}.  Joint modeling efforts require a model for intrinsic
alignments as a function of separation, redshift and galaxy properties such as luminosity, with
priors on these model parameters.  Our results suggest that the model parameters should have broad
enough priors chosen within a range selected specifically taking into account the radial weighting
of the shape measurement method used for shear estimation.  In fact, in the limit that the weak
lensing analysis is carried out using two shear estimation methods for a consistency check, it is
plausible that the best-fitting IA parameters inferred using methods that are more or less centrally-weighted may
differ.

Finally, the estimators for measuring IA should also be chosen carefully. For shape-shape correlations, $C_{11}$ is 
probably a better estimator than \wpp. In general, future studies should look at the full structure of IA in
the ($r_p,\Pi$) or ($r,\mu$) plane and tailor the estimator accordingly. \referee{Though outside the scope of our work 
in SDSS, it might also be useful for future studies to use shape measurements that probe different effective radii in a 
controlled manner, to better quantify the effects of isophotal twisting and ellipticity gradients in 
the measured galaxy alignments.}

While further investigation into multiplicative biases \referee{and ellipticity gradients} is beyond the scope of this work, 
we have
provided a first attempt at reconciling studies about intrinsic alignments in SDSS
that use different shape measurement methods.  The large differences that we uncovered suggest that
this issue is one that future weak lensing surveys cannot afford to ignore when forecasting
intrinsic alignment contamination of weak lensing signals.

\section*{Acknowledgments}
This work was supported by the National Science Foundation under
Grant. No. AST-1313169. RM was also supported by an Alfred P. Sloan Fellowship. We thank Jonathan
Blazek, Uro\v{s} Seljak, {Robert Lupton, Elisa Chisari,} and Shadab Alam for useful discussions
about this work. \referee{We thank the anonymous referee for helpful suggestions, and we} thank Ananth Tenneti for sharing the MB-II shape catalog.
	
Funding for SDSS-III has been provided by the Alfred P. Sloan Foundation, the Participating Institutions, the National Science Foundation, and the U.S. Department of Energy Office of Science. The SDSS-III web site is http://www.sdss3.org/.

SDSS-III is managed by the Astrophysical Research Consortium for the Participating Institutions of the SDSS-III Collaboration including the University of Arizona, the Brazilian Participation Group, Brookhaven National Laboratory, Carnegie Mellon University, University of Florida, the French Participation Group, the German Participation Group, Harvard University, the Instituto de Astrofisica de Canarias, the Michigan State/Notre Dame/JINA Participation Group, Johns Hopkins University, Lawrence Berkeley National Laboratory, Max Planck Institute for Astrophysics, Max Planck Institute for Extraterrestrial Physics, New Mexico State University, New York University, Ohio State University, Pennsylvania State University, University of Portsmouth, Princeton University, the Spanish Participation Group, University of Tokyo, University of Utah, Vanderbilt University, University of Virginia, University of Washington, and Yale University.

        \bibliographystyle{mnras}
        \bibliography{sukhdeep_IA_paper2,papers}		

\end{document}